\documentclass[%
 preprint,
 amsmath,amssymb,
]{revtex4-1}
\usepackage{dcolumn}
\usepackage{amsmath}
\usepackage{amssymb}
\usepackage{setspace}
\usepackage{lipsum}
\usepackage{graphicx}
\usepackage{bm}
\usepackage[T1]{fontenc}
\usepackage{color}
\usepackage{url}
\usepackage{subfigure}
\usepackage{rotating}
\usepackage{soul}
\usepackage{scalefnt}
\usepackage{hyperref}
\usepackage{scalefnt}
\usepackage{inputenc}
\setcitestyle{authoryear,open={(},close={)}}

\begin{document}

\preprint{APS/123-QED}

% \title{How are scientific works viewed?}% Force line breaks
%\title{Characterizing the Dynamics of Article Visualizations}
\title{Classification of abrupt changes along viewing profiles of scientific articles}

%with \\

\author{Ana C. M. Brito$^1$,  Filipi N. Silva$^2$, Henrique F. de Arruda$^3$, Cesar H. Comin$^4$, Diego R. Amancio$^1$ and Luciano da F. Costa$^3$}

\affiliation{$^1$Institute of Mathematics and Computer Science, University of S\~ao Paulo, S\~ao Carlos, SP, Brazil\\
$^2$Indiana University Network Science Institute, Bloomington, Indiana 47408, USA\\
$^3$S\~ao Carlos Institute of Physics,
University of S\~ao Paulo, S\~ao Carlos, SP, Brazil\\
$^4$Department of Computer Science, Federal University of S\~ao Carlos, S\~ao Carlos, SP, Brazil
}
%
%\author{Filipi N. Silva}
%\affiliation{$^2$Indiana University Network Science Institute, Bloomington, Indiana 47408, USA}%
%
%\author{Henrique F. de Arruda}
%\affiliation{$^3$S\~ao Carlos Institute of Physics,
%University of S\~ao Paulo, S\~ao Carlos, SP, Brazil}
%
%\author{Cesar H. Comin}
%\affiliation{$^4$Department of Computer Science,
%Federal University of S\~ao Carlos, S\~ao Carlos, SP, Brazil}
%
%
%\author{Diego R. Amancio}
%\affiliation{Institute of Mathematics and Computer Science\\ University of S\~ao Paulo, S\~ao Carlos, SP, Brazil\ }
%
%
%\author{Luciano da F. Costa}
%\affiliation{S\~ao Carlos Institute of Physics \\
%University of S\~ao Paulo, S\~ao Carlos, SP, Brazil\ }

\date{\today}% It is always \today, today,

\begin{abstract}
With the expansion of electronic publishing, a new dynamics of scientific articles dissemination was initiated.  Still substantially important, citations became a longer term effect.  Nowadays, many works are widely disseminated even before publication, in the form of preprints.  Another important new element concerns the views of published articles.  Thanks to the availability of respective data by some journals, such as PLoS ONE, it became possible to develop investigations on how scientific works are viewed along time, often before the first citations appear.  This provides the main theme of the present work.  More specifically, our research was motivated by preliminary observations that the view profiles along time tend to present a piecewise linear nature. A methodology was then delineated in order to identify the main segments in the view profiles, which allowed several related measurements to be derived.  In particular, we focused on the inclination and length of each subsequent segment.  Basic statistics indicated that the inclination can vary substantially along subsequent segments, while the segment lengths resulted more stable.  Complementary joint statistics analysis, considering pairwise correlations, provided further information about the properties of the views. In order to better understand the view profiles, we performed respective multivariate statistical analysis, including principal component analysis and hierarchical clustering.  The results suggest that a portion of the polygonal views are organized into clusters or groups. These groups were characterized in terms of prototypes indicating the relative increase or decrease along subsequent segments. Four respective distinct models were then developed for representing the observed segments. It was found that models incorporating joint dependencies between the properties of the segments provided the most accurate results among the considered alternatives.
\end{abstract}

\maketitle

%\tableofcontents

%\doublespacing
%\onehalfspacing
%\singlespacing

\section{\label{sec:level1}Introduction}
Science can be understood as a social activity, conceived and applied by humans. As a consequence, communication plays a critical role in scientific development, allowing important results to be disseminated and used. Interchange is important not only between scientists working on related fields, but also between those deriving results and those applying these results. In the beginnings of science, communication proceeded mostly in terms of \emph{letters} (see e.g.~\citep{peat2002certainty}), which were exchanged between scientists in order to share their most recent results. Letters gave rise to proceedings, journals and, more recently, World Wide Web-based dissemination. The study of how scientific articles are read and cited is of great importance because such knowledge provides insights about the efficiency at which science is disseminated.

Many of the existing studies in scientometrics, the area aimed at studying how science unfolds, consider citations as the main indicator of usage and interest of scientific articles. Until recently, this was one of the few available objective measurements of scientific dissemination~\citep{waltman2016review,2012three,bollen2005toward}. Yet, with the introduction of the Internet and the WWW, other statistics became available, such as the number of views, shares and downloads of articles published online. Indeed, before the Internet and the WWW, it was very difficult to count how many times a journal or article was taken from the shelves and read. The availability of these new indicators paved the way to many interesting investigations in scientometrics, motivating the new area of \emph{altmetrics}~\citep{sud2014evaluating}.

Among the new scientific indicators, the number of views has some particularly interesting features. First and foremost, it takes place at a relatively high speed, involving little delay: once published online, a work starts being viewed almost immediately. Contrariwise, the first citation of a work can take months or even years to take place. Given that views tend to be faster, they can provide insights about current trends, allowing predictions to be made. Views also tend to take place in larger numbers than citations, therefore providing a potentially more complete sample that can lead to more accurate statistical analysis. Studying views is also intrinsically important as a means to better understand its relationship with citations.
%Indeed, views are particularly important because they can be considered as \emph{precursors} to citations.
%Indeed, a much viewed article is a good candidate for being highly cited.

One of the few limitations intrinsic to views is that they provide a somewhat weaker indication of the use of the knowledge in the visualized work. Indeed, some views can be the consequence of actions of Web crawlers or surveys, without a direct implication that the reported knowledge has been somehow transferred or applied.
All in all, views-based scientometrics has potential for contributing substantially to our knowledge about how scientific information is disseminated.

%But views not only tend to be faster than citations, they are more accurate along time as a consequence of the smaller standard deviation of their onset.  Figure~\ref{f:curves} illustrates this phenomenon.  The hypothesis is that the publication of a work, marked as a dot, motivates a sigmoidal temporal profile of citations or views which is considered equal for the sake of simplicity.  The sigmoid shape accounts for the fact that, once an article is published, citations start slowly, followed by a peak of interest, and then decrease, fading out.  It is important to observe that the onset time $t_o$ for citations is much larger than for views, and also tend to have greater variability.

\begin{figure}[h]
    \centering
    \includegraphics[width=0.6\linewidth]{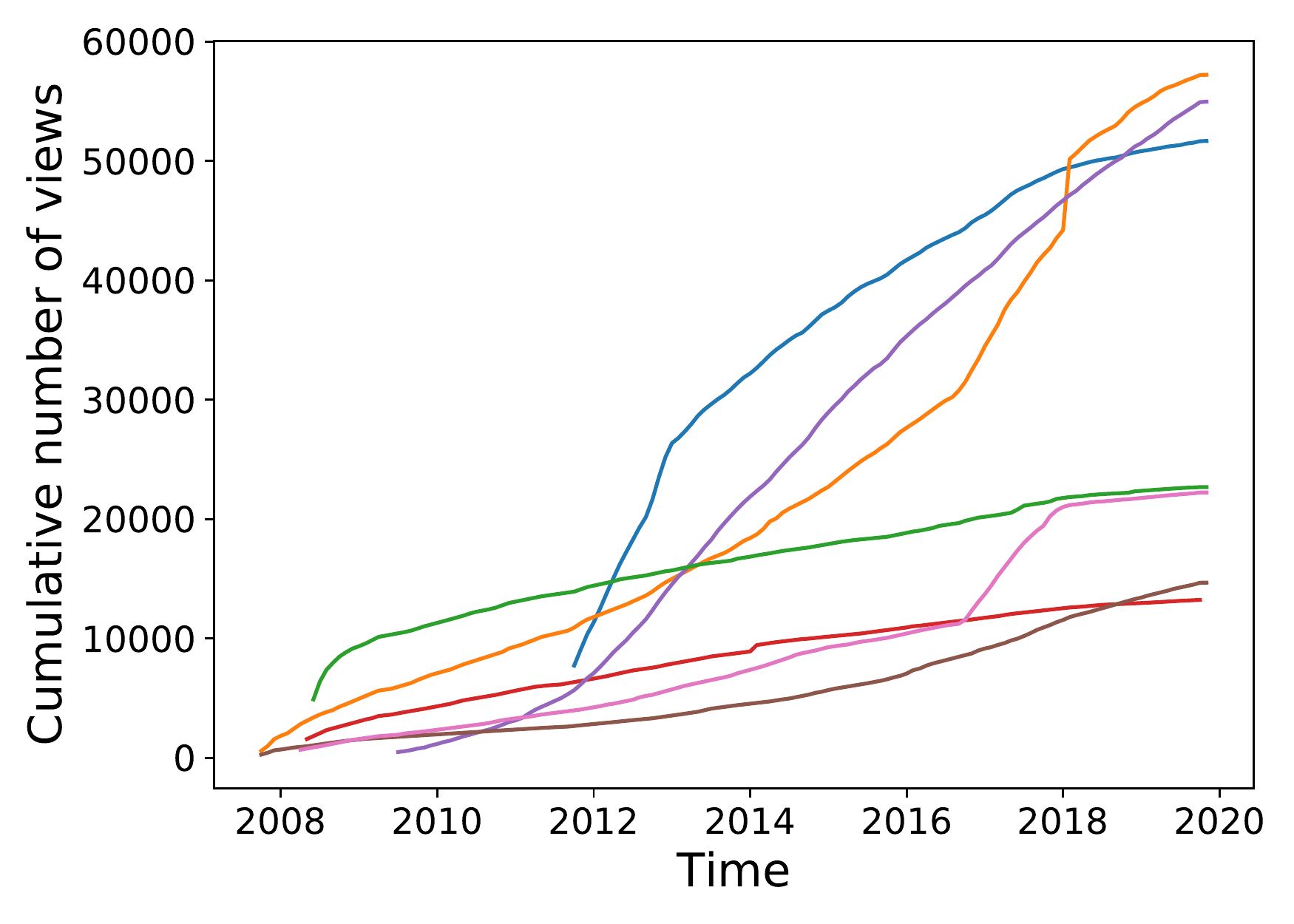} % pode ser também examples_norm.pdf
    \caption{Profiles of 6 randomly chosen PLoS ONE articles showing, in each case, the cumulative number of views along successive months. Interestingly, several of these profiles present approximate polygonal shape, with sharp turns followed by relatively straight segments.}
    \label{fig:profiles}
\end{figure}

Being mostly based to online publication, and by providing several statistics, the PLoS ONE journal~\footnote{\url{https://journals.plos.org/plosone}} represents a good resource for performing scientometric/altmetric studies focusing views as main indicators. In particular, the number of views of each article is provided along time in a month-by-month fashion. Figure \ref{fig:profiles} illustrates some view profiles for 6 randomly chosen articles.

Interestingly, some of the patters in Figure~\ref{fig:profiles} suggest a polygonal organization, including relative increases and decreases of views along subsequent segments.  This is even more interesting than the former effect, because it shows that some sort of events occur that are capable of changing substantially the visibility of a given article along time.  The study about how frequent these patterns are, and their possible organization into respective categories, constitute the main question addressed in the present work.

More precisely, viewing profiles are obtained for almost every article published in PLoS ONE. These signatures are then analysed by using state of the art numerical methods, namely segmented regression~\citep{muggeo2003estimating}, capable of fitting contiguous straight line segments along each signature, to match its original profile. The slope and transition points of these segments, corresponding to abrupt slope changes, can then be identified and used for statistical analysis and derivation of statistical models. Indeed, the current approach focuses on the derivation of these models as the means for trying to understand the properties of the considered view profiles.

The proposed models are based on variables describing the inclination angle and the extension of each subsequent linear-piecewise portion. Conditional densities are estimated from the experimental data, allowing the generation of synthetic profiles. We consider independent densities as well as Markov-1 univariate and multivariate
dependencies between variables. Then, by comparing the congruence between the original and synthetic profiles, we can estimate which of the considered models tend to be more accurate. The so-obtained model is then studied in order to identify and understand possible mechanisms producing the observed profiles.

Several interesting results are reported. First, the linear-piecewise approximation was found to be more congruent with the real-world view profiles than with synthetic profiles obtained from uniformly random events. This supports the hypothesis that the real-world profiles tend to present a polygonal structure, therefore being potentially well-represented and modeled by using this type of curve. The several parameters observed for each set of views, considering the same number of segments, revealed two relatively well-defined clusters corresponding to two main types of polygonal profiles exhibiting sequences of varying slopes. For instance, in the case of profiles containing 3 segments, one of the detected clusters is characterized by a relatively low initial slope followed by a higher slope, which is then reduced in the final segment. The other type of cluster presented opposite structure. Among the several types of statistical models considered in this work, we found that the use of conditional densities defined on joint pairs of previous parameters, more specifically by predicting the slope and interval at each current time in terms of the the previous instances considered jointly, led to the most accurate results. This suggests that each segment along the polygonal structures can be, to a good extent, estimated by the previous segment, corresponding to a memory-1 Markov dynamics that may underlies respective real-world effects.

% This article starts by presenting the adopted data as well as the concepts and methods used for respective analysis. Then, we present the results of the polygonal fitting, including a comparison with control synthetic profiles obtained from uniform statistical distribution. The analysis of the obtained features is presented next, including the identification of two main types of view profiles. Then, statistical models are developed and applied as a means to predicting and better understanding the considered profiles. The article concludes by identifying its main contributions as well as suggesting venues for future investigations.

The remaining of this paper is organized as follows. In Section~\ref{sec: related_work}, we present the related works. In Section~\ref{sec: materials}, we describe the dataset, the employed clustering algorithm, the methodology for curve characterization, and the employed statistical model. In Section~\ref{sec: results} the obtained results are discussed. Conclusions and perspectives for future works are presented in Section~\ref{sec: concluding}.

\section{Related Works}
\label{sec: related_work}
Altmetrics indicators have been employed to quantify the quality of some aspects of science~\citep{erdt2016altmetrics,sud2014evaluating,bornmann2014altmetrics,galligan2013altmetrics}. Some analyses account for the importance of articles~\citep{huang2018altmetric} and others for measuring characteristics of scientists~\citep{lariviere2013bibliometrics,ioannidis2014bibliometrics}. Interestingly, many altimetric measurements are not found to be correlated to citation counts~\citep{erdt2016altmetrics}, which indicates that these metrics can measure different and complementary information.

The Altimetric Attention Score (AAS) has been used to measure the importance of papers~\citep{huang2018altmetric}. This index measure mentions of papers in social media (e.g., Twitter and Facebook). \cite{huang2018altmetric} found correlations between AAS and the number of citations for some journals. More specifically, by considering papers obtained from PLoS journals, the authors measured the Spearman correlation between a normalized AAS and a normalized count of citations. Interestingly, in the case of Medicine articles, this correlation was not found. In another study that considered highly cited papers, correlations were found in the comparison between metrics obtained from social networks and the number of citations~\citep{thelwall2013altmetrics}.

Another important information that can be measured is the difference between the behavior of new and old articles. Due to the fast evolution of computer-related areas, computer science researchers attribute great importance to conference papers. By considering papers of journals and conferences, \cite{thelwall2019mendeley} compared the number of citations with \emph{Mendeley Readership}. Mendeley Readership measures the number of users that include a given article to their account. They found that the number of Mendeley reads and the number of citations is correlated for both journals and conferences. However, in the case of old conference papers, a similar correlation was not found.  Furthermore, \cite{schlogl2014comparison} compared citations, downloads, and Mendeley Readership of two information systems journals, and found high Spearman correlations when comparing downloads with citation and downloads with Mendeley Readership. However, in the comparison between Mendeley Readership with citations, a moderate correlation was found.

One particularly important measurement is the number of views, for which some distinct aspects have been analyzed. For instance, the number of views can be linearly correlated with the age of a paper~\citep{priem2012altmetrics}, in which older articles tend to have more views. Furthermore, in~\citep{de2015relationship}, the authors investigated the relationship between the number of article's views and the mentions of articles in \emph{Twitter}. More specifically, their study suggests that views obtained from \emph{tweets} are not related to views and citations. Other scholars also analyze the number of article views and downloads and found that the latter is much more correlated with citations than views~\citep{wang2014attention}. In an analysis regarding medical papers, the documents with high early views counts tend to be more cited than others~\citep{perneger2004relation}. In a comparison among different altmetrics extracted from \emph{PlumX}, the measurements of views and downloads were found to have the most extended life cycles~\citep{ortega2018life}. Their results also indicate that mentions in Twitter and blogs can impact the number of views and downloads.

Researchers have also been comparing the number of downloads with citations~\citep{erdt2016altmetrics}. In some studies, a correlation between the number of downloads and citations was found~\citep{brody2006earlier,watson2009comparing,perneger2004relation,jamali2011article}. A form of fast transmitting scientific results is publishing papers on \emph{arXiv}, which is a preprint repository. In a study that took into account early citations~\citep{shuai2012scientific}, the authors compared \emph{arXiv} downloads and Twitter mentions and found that there is a correlation between tweets and downloads. By considering a given journal, in~\cite{moed2005statistical}, the authors found that in the following three months after a citation, the number of downloads tends to increase. Additionally, for the same dataset, when downloads and citation distributions are compared, the older, the more similar~\citep{moed2005statistical}.

Another possibility of analysis is the comparison between altmetric indicators and the linguistic characteristics of the papers. In~\cite{chen2020exploring}, many distinct features were measured from papers of PLoS, which included text length, lexical diversity, lexical density, among others. Interestingly, for the majority of the considered characteristics, no significant correlation was found. However, for some PLoS journals, the title lengths and average sentence length were found to play an important role in the number of views and downloads. This correlation was also found for~\cite{jamali2011article}. However, \cite{duan2017download} did not find a correlation between the total number of downloads and title lengths.

\section{Materials and Methods}
\label{sec: materials}

\subsection{The Dataset}
We extracted information from all the papers published in PLoS ONE up to 2016. For that, we employed a semi-automatic extraction of paper metadata using ALM API. This data was collected in November 2019. For each article, the dataset contains information about the number of views per month, as well as other social media features, including the number of tweets and shares. The latest properties still require some post-processing and further validation before use. Thus, we only focus on analyzing the number of views along time, which resulted in a total of $162,534$ view profiles. The total number of publications views is divided into HTML, PDF, and XML. In order to be compatible with the way in which views are understood and exhibited in the PLoS ONE web site, we used the sum of all types, instead of only the HTML or PDF views.

%  Este é o trecho que escrevi sobre views a partir de outros sites.  Luciano
It is important to observe that the consideration of data like in the PLoS ONE dataset will typically not account for views through other means such as pre-prints, department reports, etc.  However, these views are expected to be relatively less frequent than to those recorded in the adopted database.

% Apart from the data obtained from Plos ONE, we also estimated the citations from Microsoft Academic Graph (MAG)~\cite{sinha2015overview}. This dataset was obtained on June 25, 2020, and comprises many different types of records: book, book chapter, conference and journal papers, dataset, patent, repository, and unknown. Furthermore, for each document, other information is also provided, which includes document citations and publication date. In order to obtain the time series of the citations, we counted the cited papers that have been published within each month.

%\textcolor{blue}{
%INFORMACOES SOBRE OS DADOS:
%Quantidade original de curvas na base: 162534;
%Numero de curvas depois de aplicado o algoritmo de segmentacao (sem tirar outliers): 129845;
%Curvas com 2 intervalos: 177 0.15\%;
%Curvas com 3 intervalos: 5263 4.5\%;
%Curvas com 4 intervalos: 30883 26.37\%;
%Curvas com 5 intervalos: 80796 68.98\%
%}

\subsection{Agglomerative clustering}

In order to identify groups of related articles regarding their viewing patterns, agglomerative clustering~\citep{mullner2011modern,jain1999data} will be applied in the experiments. An important choice concerns the linkage criterium to be used for aggregating datapoints, which often consists in adopting Ward's approach.  One disadvantage of this method is that it tends to identify groups even when the groups are not well-defined or do not exist, which are henceforth referred as false positives. The linkage method that is likely most suited for avoiding false positives is the \emph{single-linkage}~\citep{tokuda2020revisiting}. This simple criterium, which links two groups based on the smallest distance between any two objects in those groups, only leads to the detection of clusters if there is a marked separation between the points belonging to the clusters. For instance, for uniformly distributed data the single-linkage approach will result in a dendrogram containing similar cophenetic distances~\citep{sokal1962comparison}, thus indicating that clusters cannot be derived from the dendrogram. On the other hand, other common linkage criteria, in particular the Ward's method, tend to result in dendrograms indicating clusters in the data.

The robustness of the single-linkage method to false positives has an important consequence. Clusters found by this method are unlikely to be caused by statistical fluctuation, missing data or noise. Thus, the identified clusters can be considered as being more statistically significant.

\subsection{Curve characterization} \label{sec:cesar}
\begin{figure}[h]
    \centering
    \includegraphics[width=0.55\textwidth]{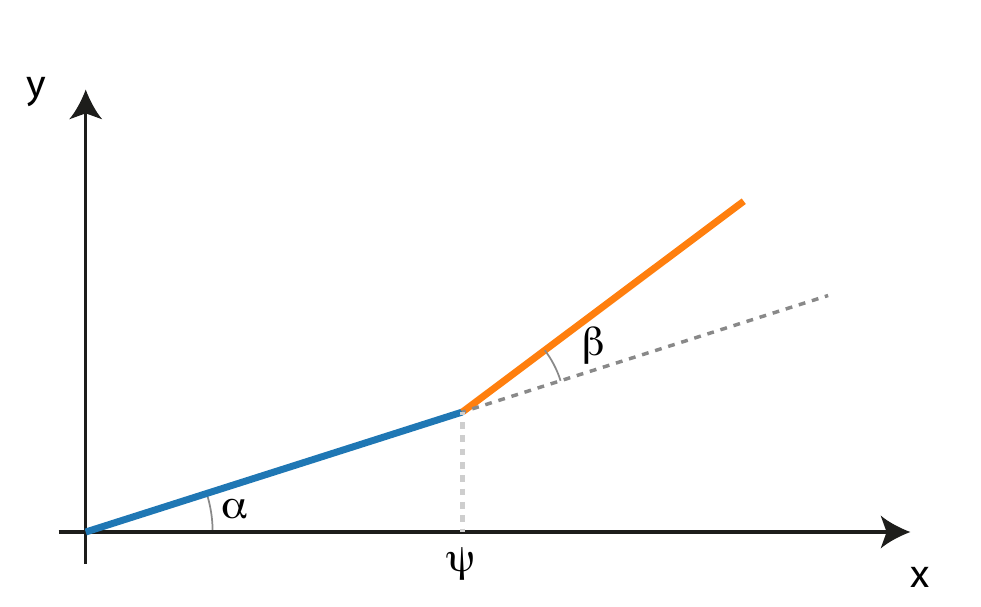}
    \caption{Example of segmented line with one breakpoint. %\textcolor{red}{??Acho que a interpretacao que dei para o parametro $\gamma$ esta incorreta. Ele esta relacionado com a distancia entre as curvas, mas nao pode ser a distancia indicada. Se ninguem conseguir interpretar o parametro, melhor remover a figura b} Tirei figura b
    }
    \label{fig:methodexample}
\end{figure}

Following~\citep{muggeo2003estimating}, a breakpoint is modeled as two straight lines joined at point $\psi$, that is,
\begin{equation}
    y = \alpha x + \beta (x-\psi)_+ \label{eq:break_model}
\end{equation}
where $(x-\psi)_+=(x-\psi) I(x>\psi)$, with $I(A)=1$ if condition $A$ is true and $I(A)=0$ otherwise. $\psi$ is the position of the breakpoint, $\alpha$ is the slope of the line before the breakpoint and $\alpha+\beta$ is the slope after the breakpoint. Therefore, $\beta$ is the difference in slopes between the two lines. This is illustrated in Figure~\ref{fig:methodexample}.
The term $(x-\psi)_+$ can be rewritten as
\begin{equation}
    (x-\psi)_+= (x-\psi^s)_+ + (\psi-\psi^s)(-1)I(x>\psi^s)
\end{equation}
which represents a first order Taylor expansion of $(x-\psi)_+$ at $\psi_s$. By defining $U^s=(x-\psi^s)_+$ and $V^s=-I(x>\psi^s)$, Equation~\ref{eq:break_model} can be rewritten as
\begin{equation}
    y = \alpha x + \beta U^s + \gamma V^s \label{eq:break_model2}.
\end{equation}
Thus, the piecewise linear curve is represented as a linear combination between variables $x$, $U^s$ and $V^s$.  Representing as $x_1,x_2,\dots,x_n$ and $y_1,y_2,\dots,y_n$ the values measured for, respectively, the $x$ and $y$ variables, we want to find $\beta$, $\gamma$ and $\psi_s$ which minimizes the sum of squared residuals for the model in Equation~\ref{eq:break_model2}. \cite{muggeo2003estimating} showed that this can be done according to the following procedure:
\begin{enumerate}
    \item Set an initial value for $\psi^s$,
    \item Calculate $U^s=(x_i-\psi^s)_+$ and $V^s=-I(x_i>\psi^s)$ for each data point $i$,
    \item Fit the model in Equation~\ref{eq:break_model2} using linear regression,
    \item Improve the breakpoint estimation according to $\psi^{s+1}=\gamma/\beta + \psi^s$,
    \item Repeat 1-4 until convergence.
\end{enumerate}

In case of multiple breakpoints, the process is similar. The breakpoints are modeled as

\begin{equation}
    y = \alpha x + \sum_{k=1}^{N}\beta_k (x-\psi_k)_+ \label{eq:break_model_mult}
\end{equation}
where $N$ is the number of breakpoints. $\beta_k$ represents the difference in slopes between segments $k$ and $k+1$. Equation~\ref{eq:break_model_mult} can be rewritten as

\begin{equation}
    y = \alpha x + \sum_{k=1}^{N}\beta_k U_k^s + \sum_{k=1}^{N}\gamma_k V_k^s\label{eq:break_model_mult2}.
\end{equation}
The parameters of Equation~\ref{eq:break_model_mult2} are found by least squares regression, and the breakpoints estimates are updated as $\psi^{s+1}_k=\gamma_k/\beta_k + \psi^s_k$.

The segmented regression requires the number of breakpoints to be known a priori. \cite{muggeo2010efficient} defined a procedure for finding an appropriate number of breakpoints. First, the aforementioned segmented regression method is applied using a large number of candidate breakpoints. Then, during the optimization procedure a breakpoint is removed if $\beta_k\approx 0$, which indicates that the slopes of segments $k$ and $k+1$ are too similar, and if two breakpoints are too close to each other. Next, breakpoints that do not significantly contribute to the residual are removed using the least-angle regression algorithm~\citep{efron2004least}.

For our analysis, the segmented regression was applied using the \emph{segmented} package in the R language. The first point of each viewing profile was removed since it represents the number of article views along the first calendar month when the article was published. Thus, if the article was published at the end of the month, it will have an unreasonably low number of views for that month.

\subsection{Statistical Model} \label{sec:statistical}
To better understand the linear-piecewise nature observed in the studied view profiles, we propose a set of simple statistical models. Each of the models progressively incorporate more information about the data. In particular, it takes into account more relationships among parameters recovered from approximating the view profiles by using the segmented regression algorithm.

We start the analysis by applying the segmentation algorithm to all the curves, resulting in a set of piecewise linear curves. Each curve is given in terms of the breakpoints and slopes that estimates the original view profile. By considering only the curves with a fixed number $N$ of segments, we extract the piecewise parameters given by the algorithm: $l_i~(1 \leq i \leq N)$, representing the length of the $i^{th}$ segment between two breakpoints, and $\alpha_i~(1 \leq i \leq N)$, the inclination of the $i^{th}$ segment relative to the x-axis.

One of our goals is to produce synthetic profiles according to statistical models providing support to understand the real-world profiles. Here, we propose four types of models: null model, independent distribution model, Markov-1 univariate, and multivariate models.

For the models, we consider that $l_i$ and $\alpha_i$ are random variables given by probabilities $P(\alpha_{i+1} | \alpha_i)$, $P(l_{i+1}|l_i)$, $P(l_{i+1}| l_i, \alpha_i)$, and $P(\alpha_{i+1}|l_i,\alpha_i)$ and employ the conditional probability equation, as follows
\begin{equation}
    P(B|A) = \frac{P(A \cap B)}{P(A)}.
    \label{eq:condprob}
\end{equation}
% Conditional probability (Equation \ref{eq:condprob}) is used to explain dependencies between the random variables: $P(\alpha_{i+1} | \alpha_i)$, $P(l_{i+1}|l_i)$, $P(l_{i+1}| l_i, \alpha_i)$, and $P(\alpha_{i+1}|l_i,\alpha_i)$.
% \begin{equation}
%     P(B|A) = \frac{P(A \cap B)}{P(A)}
%     \label{eq:condprob}
% \end{equation}

In our simplest model, which we call null model, $\alpha_i$ and $l_i$ are generated following uniform distributions. Each $\alpha_i$ is drawn between $[0,90)$ and the $l_i$ in the range $[0,1)$. The independent distribution model takes into account independent densities. For samples obtained for a variable $X$, the probability density function (PDF) of the variable is estimated using kernel density estimation with a Gaussian kernel. The PDFs of the $\alpha_i$ and $l_i$ variables were estimated from the articles data and used for generating synthetic values.

In the \emph{Markov-1 univariate model}, given a sequential random variables $X_i$, the values of the next state $i+1$ are drawn from the univariate conditional probabilities given by the current state $X_{i}$.
%In such a sense, the model has a $1$ level ``memory''\textcolor{red}{?? Esse modelo nao eh um processo de Markov usual? Não sei se é comum chamar de memória 1}. FIXED
More specifically, starting from an initial independent distribution of $X_1$, the first state is generated. The next state is drawn from $P(X_{i+1}|X_i)$. To estimate $P(X_{i+1}|X_i)$, both the $\alpha_i$ and $l_i$ parameters obtained for the data were partitioned into bins.

Finally, the Markov-1 multivariate model uses the combined joint probabilities for $\alpha_i$ and $l_i$ to calculate $\alpha_{i+1}$ and $l_{i+1}$. More specifically, the initial independent joint distribution $P(\alpha_1,l_1)$ is used to generate the first states ($\alpha_1$ and $l_1$ values), then the distributions $P(\alpha_{i+1}|\alpha_i,l_i)$ and $P(l_{i+1}|\alpha_i,l_i)$ are employed to calculate the next corresponding states.

%Some other models were tested, combining Markov-1 univariate for $\alpha$ and independent distribution model for $l$ (the opposite was also verified). But the main results were obtained from the four models explained above.

%\textcolor{red}{Adicionar figura com variáveis da modelagem estatística. $\alpha_1$ $\alpha_2$, $l_1$, $l_2$, $l_3$}

\section{Results and Discussion}
\label{sec: results}

In this Section, we discuss the adherence of the real data to segmented lines in Section~\ref{sec:a}. The basics statistics of the real data are discussed in Section~\ref{sec:b}. In Section~\ref{sec:c}, we analyze the correlation between the parameters of the segmented lines describing the evolution of paper views. In Section~\ref{sec:d}, we present a discussion on the clustering behavior of the segmented view profiles. Finally, in Section~\ref{sec:e}, we provide an evaluation of the model aimed at reproducing the behavior of paper views along time.

% remoção de outliers e artigos com menos de 3 anos de publicação (definir o que é lifetime)
%
%The results are organized....???

\subsection{Segmented regression adherence} \label{sec:a}
\begin{figure}
    \centering
    \includegraphics[width=0.9\textwidth]{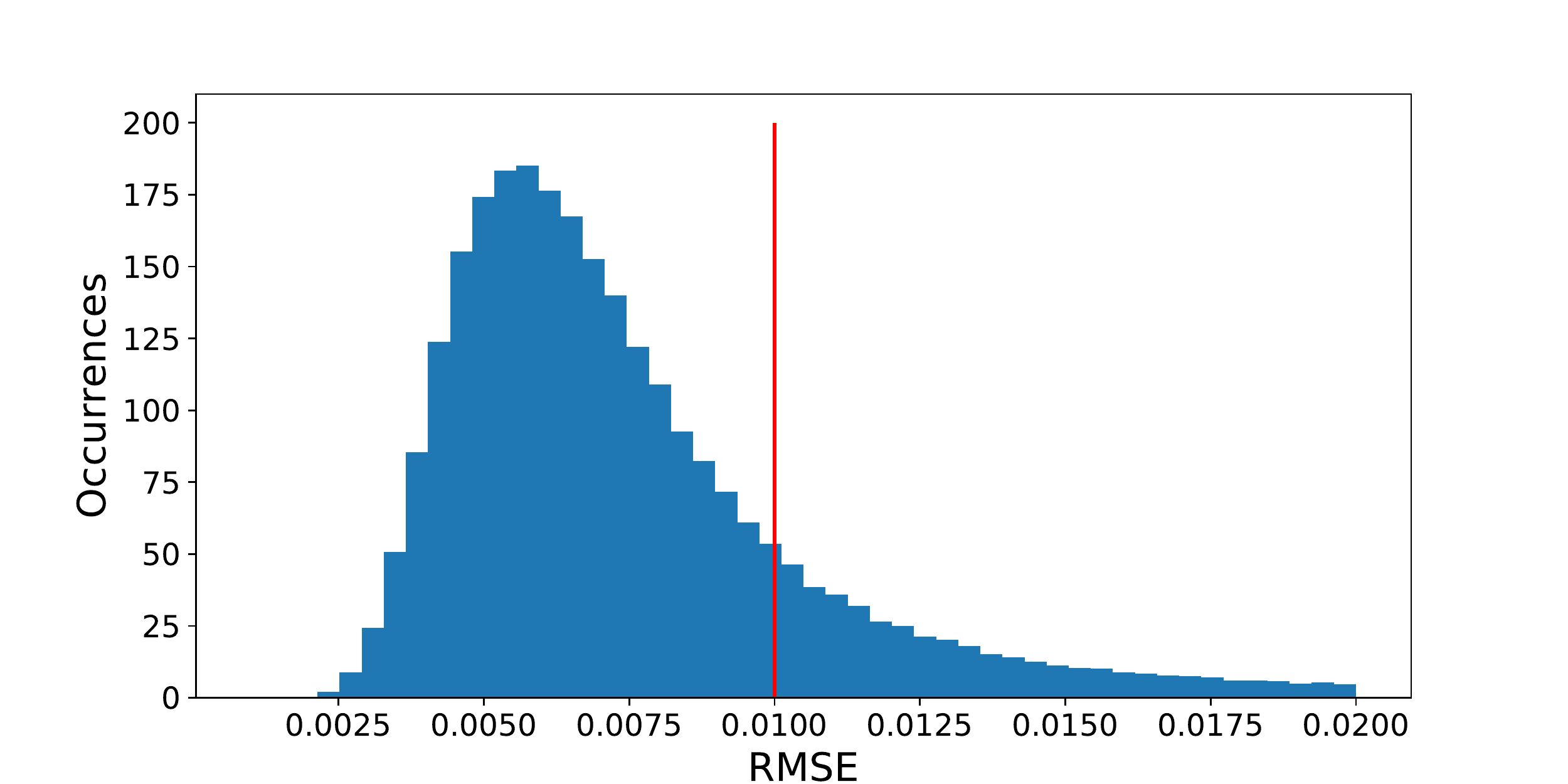}
    \caption{RMSE distribution for the segmented regressions of the view profiles. The red line indicates the threshold RMSE value; only the curves on the left were selected to be used in the experiments. With the established threshold, roughly 80\% of all view curves were analyzed.}
    \label{fig:msedist}
\end{figure}

The first step of the analysis is the application of the \textit{segmented regression} algorithm to all the obtained view profiles. {In order to do so, each view profile was normalized to the range $[0,1]$ in both the time and number of views axes. Therefore, the cumulative number of views of an article for the most recent month in the dataset (September of 2019) is always 1.} A preliminary analysis by visual inspection suggested that most of the profiles contain $5$ or less linear segments. In order to check which curves best adhere to a structure of segmented lines, we computed the Root Mean Square Error (RMSE)~\citep{hyndman2006another} between the curves derived from the algorithm and the original data. Figure~\ref{fig:msedist} shows the distribution of the RMSE for the considered view profiles. %All visualization curves with $\textrm{RMSE} \geq 0.10$ were disregarded from our analysis. %[waiting for ??? RMSE new plot to comment].

\begin{figure}
    \centering
    \subfigure[Example of a valid profile for \textit{segmented} regression (RMSE = 0.006).]{\includegraphics[width=0.4\textwidth]{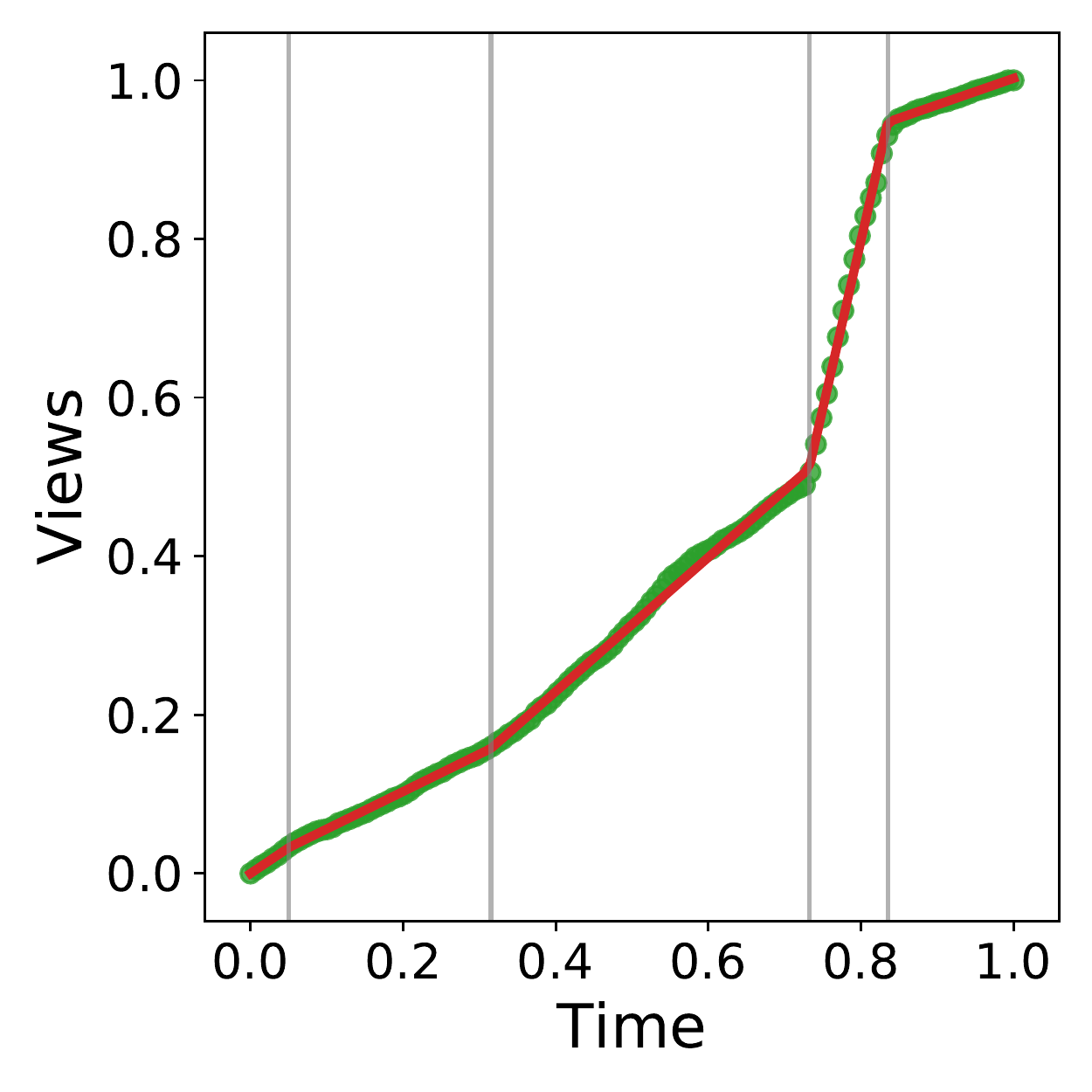}} %10.1371/journal.pone.0001565
    \subfigure[Example of a invalid profile for segmented regression (RMSE = 0.0142).]{\includegraphics[width=0.4\textwidth]{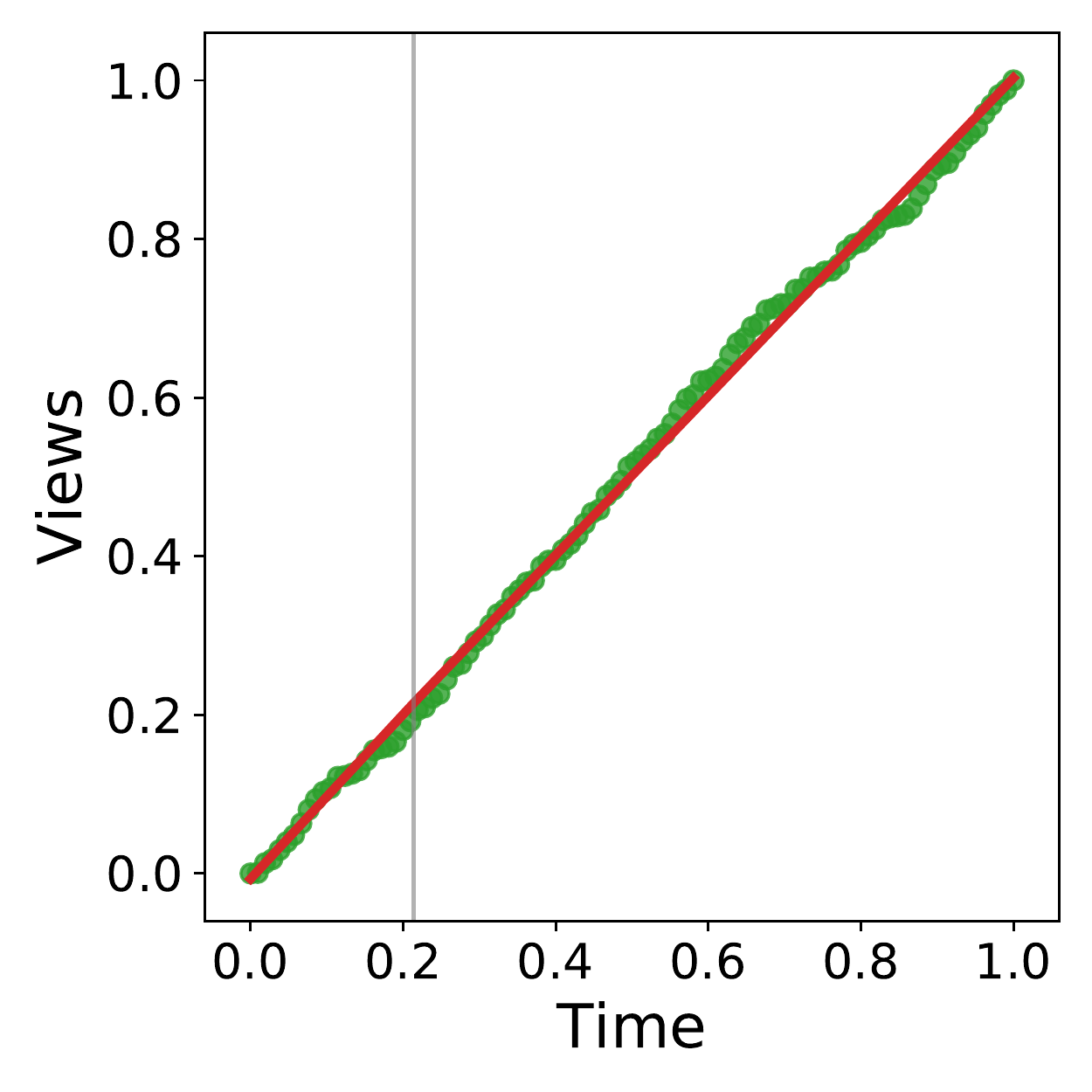}} %10.1371/journal.pone.0000341
    \caption{Examples of view profiles and corresponding linear-piecewise curves. The \textit{segmented} regression result is the red lines, the original values are the green points, and the breakpoints occur where there are the gray vertical lines.
    }
    \label{fig:validinvalidexamples}
\end{figure}

To test if the curves adhere to the segmented regression model, we chose a conservative threshold based on the obtained RMSE values. More specifically, we selected only the curves for which the regression resulted in an RMSE value lower than $0.01$ (shown as a red line in Figure~\ref{fig:msedist}). Figure~\ref{fig:validinvalidexamples} shows two examples of views profiles (green dots) -- characterized by a higher (a) and lower (b) adherence to the model -- and their respective piecewise curves (red segments) together with the breakpoints (gray vertical lines). About $80\%$ of the curves have been found to pass the RMSE test and were selected for further analyses. %Table \ref{tab:mseoriginal} summarizes the number of selected and discarded curves according to their adherence to the model.
%\textcolor{blue}{(REMOVI A TABELA DO RMSE)}

%\begin{table}[]
 %   \centering
%    \caption{Segmented regression adherence in original profiles.}
%    \begin{tabular}{|r|l|l|}
%        \hline
%        & \textbf{Frequency} & \textbf{Percent} \\
%        \hline
%        \textbf{RMSE < 0.01} & 129872 & 79.90\\
%        \hline
%        \textbf{RMSE} $\geq$ \textbf{0.01} & 31995 & 19.69\\
 %       \hline
%        \textbf{Error} & 667 & 0.41\\
%        \hline
%        \textbf{Total} & 162534 &100\\
%        \hline
%    \end{tabular}
%    \label{tab:mseoriginal}
%\end{table}{}

In order to validate the hypothesis that real-world view profiles tend to present a polygonal structure, control synthetic profiles were generated. For each original profile, a synthetic one with the same number of points was created. Synthetic views were generated following a discrete uniform distribution $[0,h_M)$ for the number of views an article received in a month, where $h_{M}$ is the largest number of views among all articles in the real data.
%
%Table \ref{tab:mseonullmodel} shows
We found a similar adherence compared to the original profiles when the \textit{segmented} method was applied with the same parameters in synthetic profiles: about 78\% of the lines passed the RMSE test. Despite this, there is a significant difference between the angle distribution in original and synthetic piecewise curves (see Figure~\ref{fig:angles} of the Supplementary Information (SI)). The latter is centered on 45-degree values. %It means that synthetic profiles have, in general, a diagonal line shape.
The differences between the above mentioned synthetic profiles and the real curves mean that the observed view curves for PloS ONE articles cannot be explained by this simple stochastic process.
%\textcolor{red}{PRECISA FALAR ALGO SOBRE AS FIGURAS \ref{fig:angles} E \ref{fig:intervals}, ELAS NAO FORAM MENCIONADAS NO TEXTO!!! (estao NO MATERIAL SUPLEMENTAR agora)}

%\begin{table}[]
%    \centering
%    \caption{Segmented regression adherence in control synthetic profiles.}
%    \begin{tabular}{|r|l|l|}
%        \hline
%        & \textbf{Frequency} & \textbf{Percent} \\
%        \hline
%        \textbf{RMSE < 0.01} & 127344 & 78.35\\
%        \hline
%        \textbf{RMSE} $\geq$ \textbf{0.01} & 34079 & 20.96\\
%        \hline
%        \textbf{Error} & 1094 & 0.67 \\
%        \hline
%        \textbf{Total} & 162517* &100\\
%        \hline
%    \end{tabular}
%    \label{tab:mseonullmodel}
%\end{table}{}

% \begin{figure}
%     \centering
%     \includegraphics[width=0.8\textwidth]{imgs/rmse_by_seg.pdf}
%     \caption{RMSE distribution for the segmented regressions by group of profiles with the same number of segments.}
%     \label{fig:mse_segs}
% \end{figure}
%????

\subsection{Basic Statistics} \label{sec:b}

\begin{figure}
    \centering
    \subfigure[][Lifetime distribution]{\includegraphics[width=0.45\textwidth]{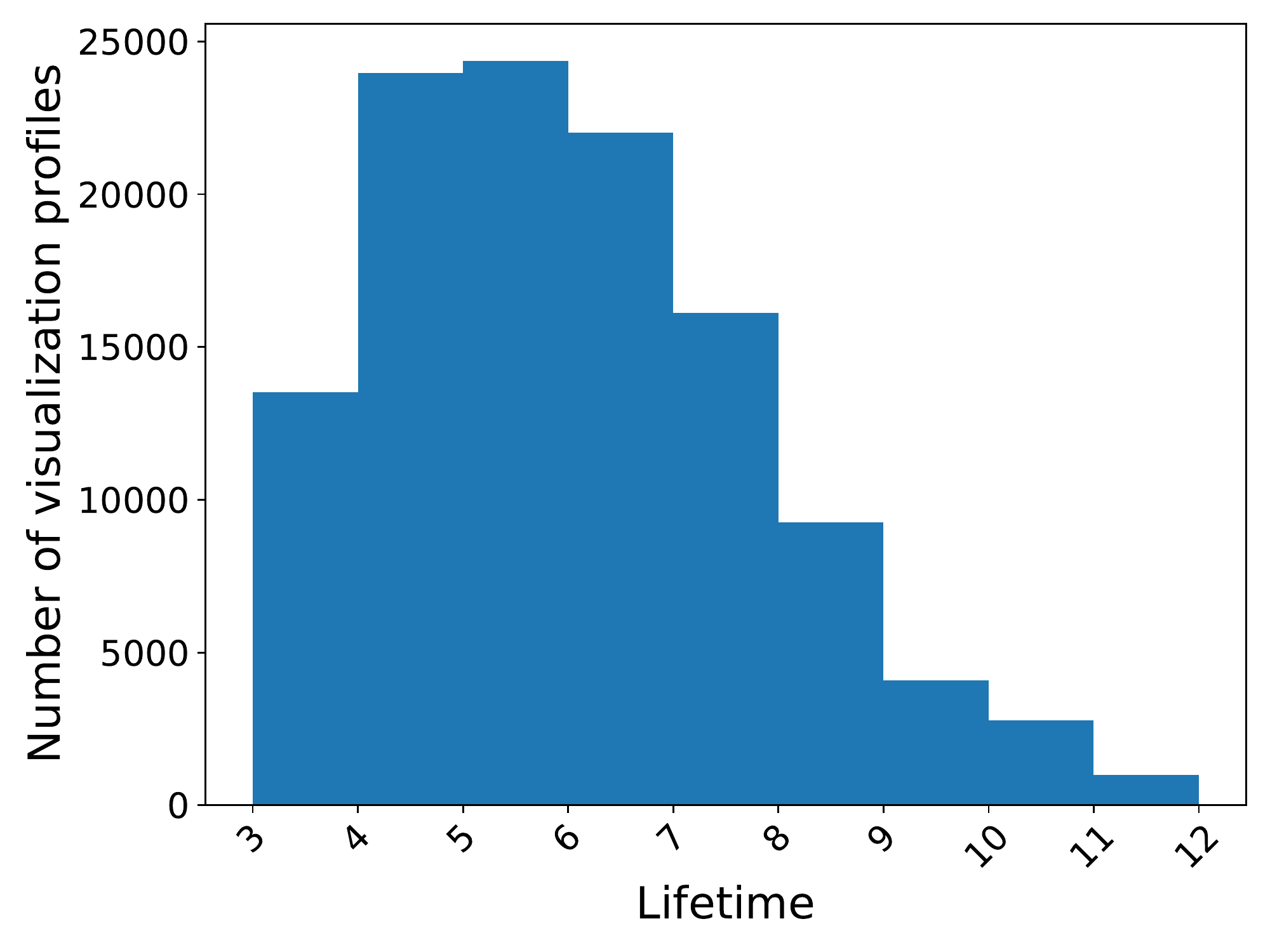}\label{fig:lifetime}}
    \subfigure[][Views distribution]{\includegraphics[width=0.45\textwidth]{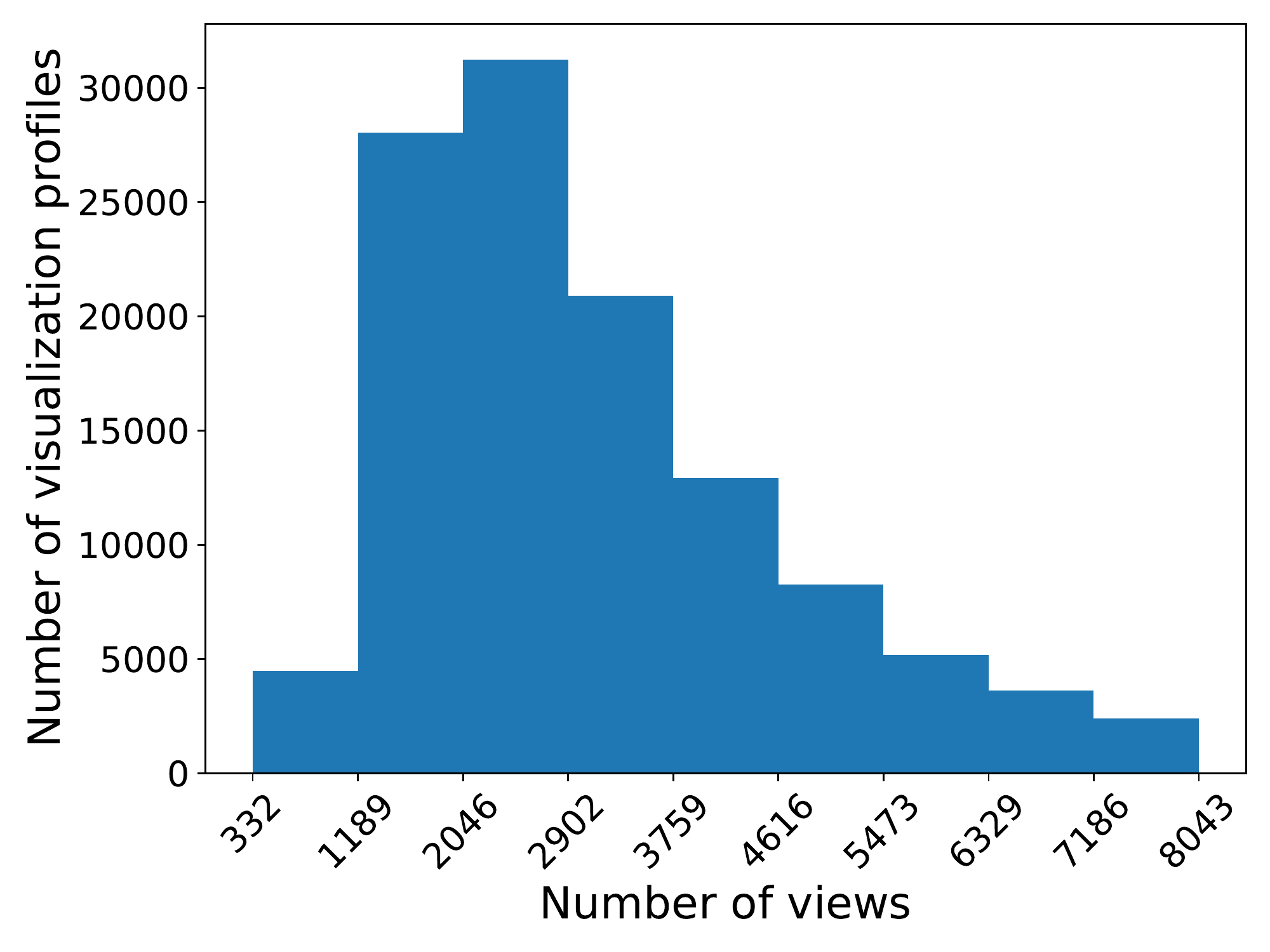}\label{fig:visual}}
    \caption{Distributions of (a) lifetime (in years) and (b) number of views of the selected view profiles (after removing outliers). From a given profile, lifetime is the period since its publication to the moment data was collected, and views refer to the total paper views along its lifetime.
    }
    \label{fig:dists}
\end{figure}

We analyzed some basic statistics of the considered views curves, i.e., those that passed the segmented regression test described in Sections~\ref{sec:cesar} and~\ref{sec:a}. Figure~\ref{fig:dists} shows the lifetime and cumulative views distributions of the views curves. The lifetime is the number of years since a certain article was published (up to 2016), and ranges predominantly between $4$ and $7$ years according to Figure \ref{fig:lifetime}. Also, most papers have from $1,500$ to $3,500$ views, as shown in Figure \ref{fig:visual}.
%As expected, we also found a moderate correlation between lifetime and number of views \textcolor{black}{(see Figure \ref{fig:bivariate} of the SI)}.

%---------------------------------------------
% Na Figura \ref{fig:segs} eh apresentada a distribuicao de numero de segmentos obtidos por visualization profiles. O melhor ajuste de numero de segmentos foi definido pelo proprio algoritmo de segmentacao, para os dados utilizados, a maior parte foi ajustado com 5 segmentos (limiar maximo pre-definido).

Most of the curves passing the RMSE test have been found to be modeled best by $5$ linear segments. The proportion of curves found to be
described by 2, 3, 4 and 5 intervals are: 0.1\%, 4.5\%, 26.4\% and 69.0\%, respectively.
It is not a surprise since the segmentation algorithm chooses the best number of segments, and the tendency is to have more segments to better adhere to the regression and the original data.
We also found that the lifetime average (and standard deviation) seems to increase with the number of segments of the curves, this is shown in Figure~\ref{fig:lifetimebysegments} of the SI.  This effect indicates that, as could be expected, the intricacy of the views profiles tend to increase with time.% std aumenta mas a quantidade de profiles aumenta muito mais ????

% \begin{figure}
%     \centering
%     \includegraphics[width=0.45\textwidth]{imgs/segments.pdf}
%     \caption{Distribution of the number of segments identified in the visualization profiles after applying the \textit{segmented} method.}
%     \label{fig:segs}
% \end{figure}

From the segmented patterns, we extracted the parameters describing the curves according to the proposed segmentation method, more specifically, the angles $\alpha_i$ and segment lengths $l_i$. The respective parameter distributions are shown in Figure~\ref{fig:histbyvar} of the SI. We found that the angles tend to become smaller along the consecutive segments, possibly reflecting the loss of visibility with time. The lengths of segments distributions become narrower with time, meaning that the monthly views of articles tend to change more rapidly as the article gets older.

%---------------------------------------------

\subsection{Joint Distributions} \label{sec:c}
An important step to understanding how subsequent segments are related is the analysis of the joint distributions among their parameters. In particular, we focus our analysis on the bivariate distributions of $\alpha_i$ and $l_i$ for subsequent segments.

\begin{figure*}
\centering
\subfigure[$\rho$ = -0.54]{\includegraphics[width=7cm]{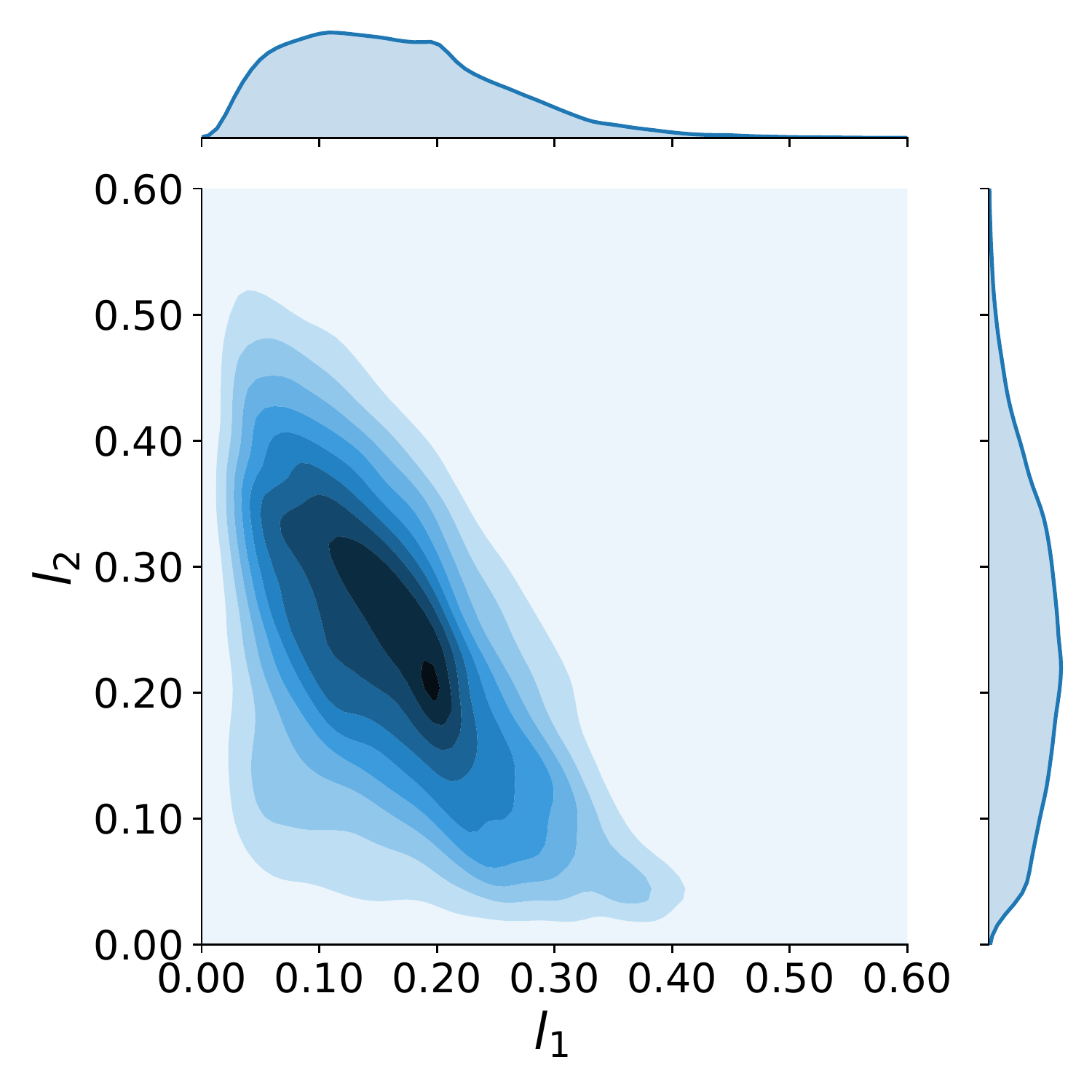}}
\subfigure[$\rho$ = -0.47]{\includegraphics[width=7cm]{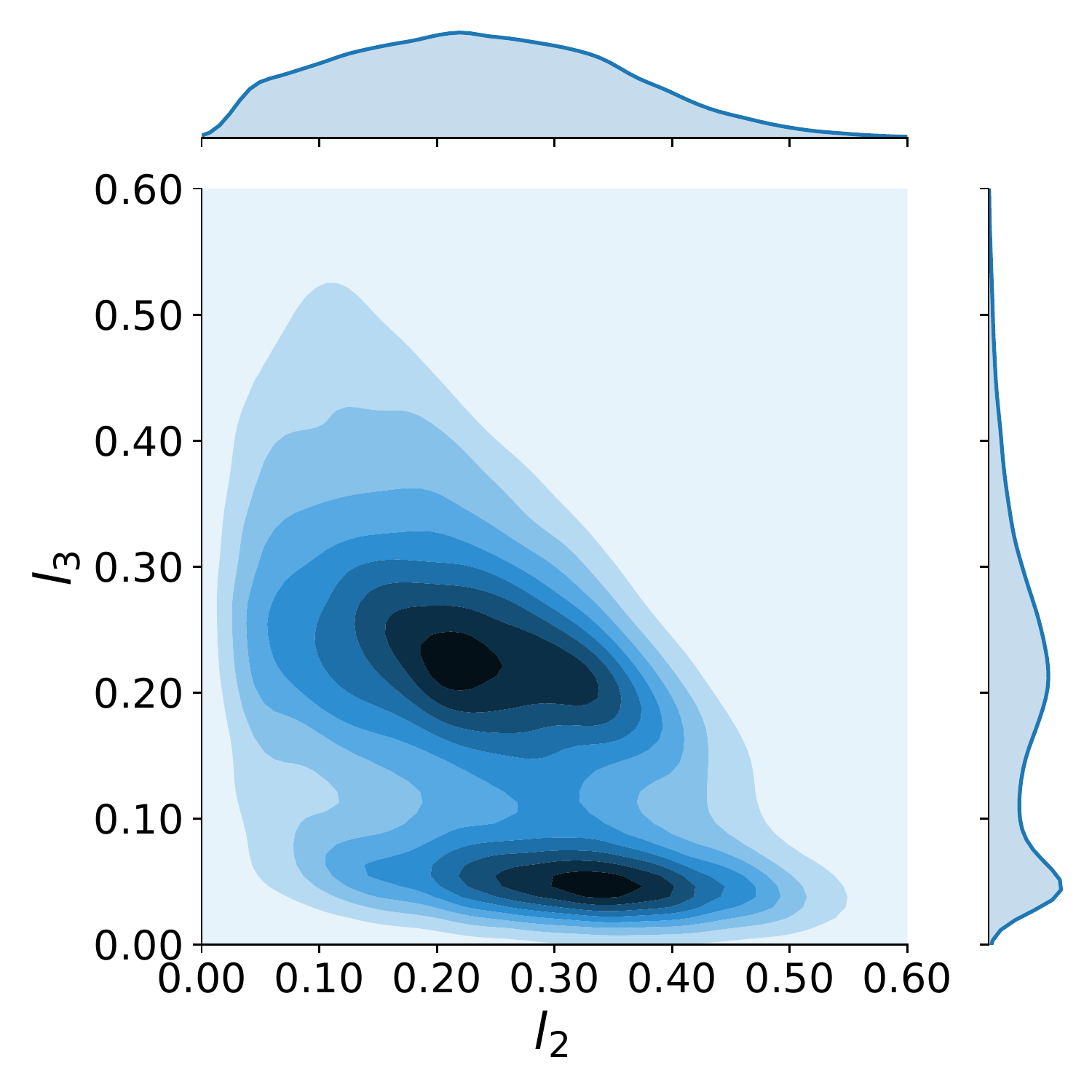}}
\subfigure[$\rho$ = -0.46]{\includegraphics[width=7cm]{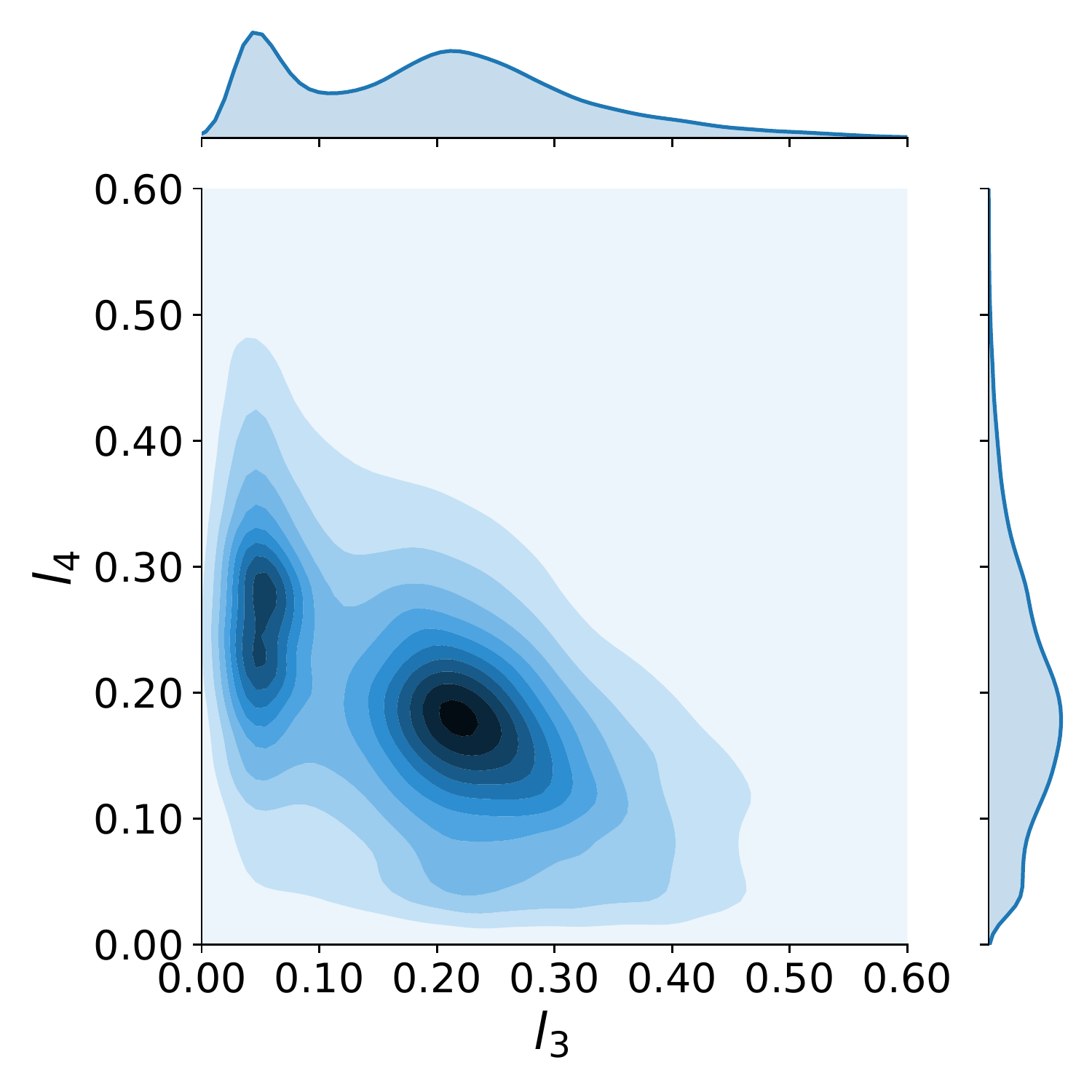}}
\subfigure[$\rho$ = -0.44]{\includegraphics[width=7cm]{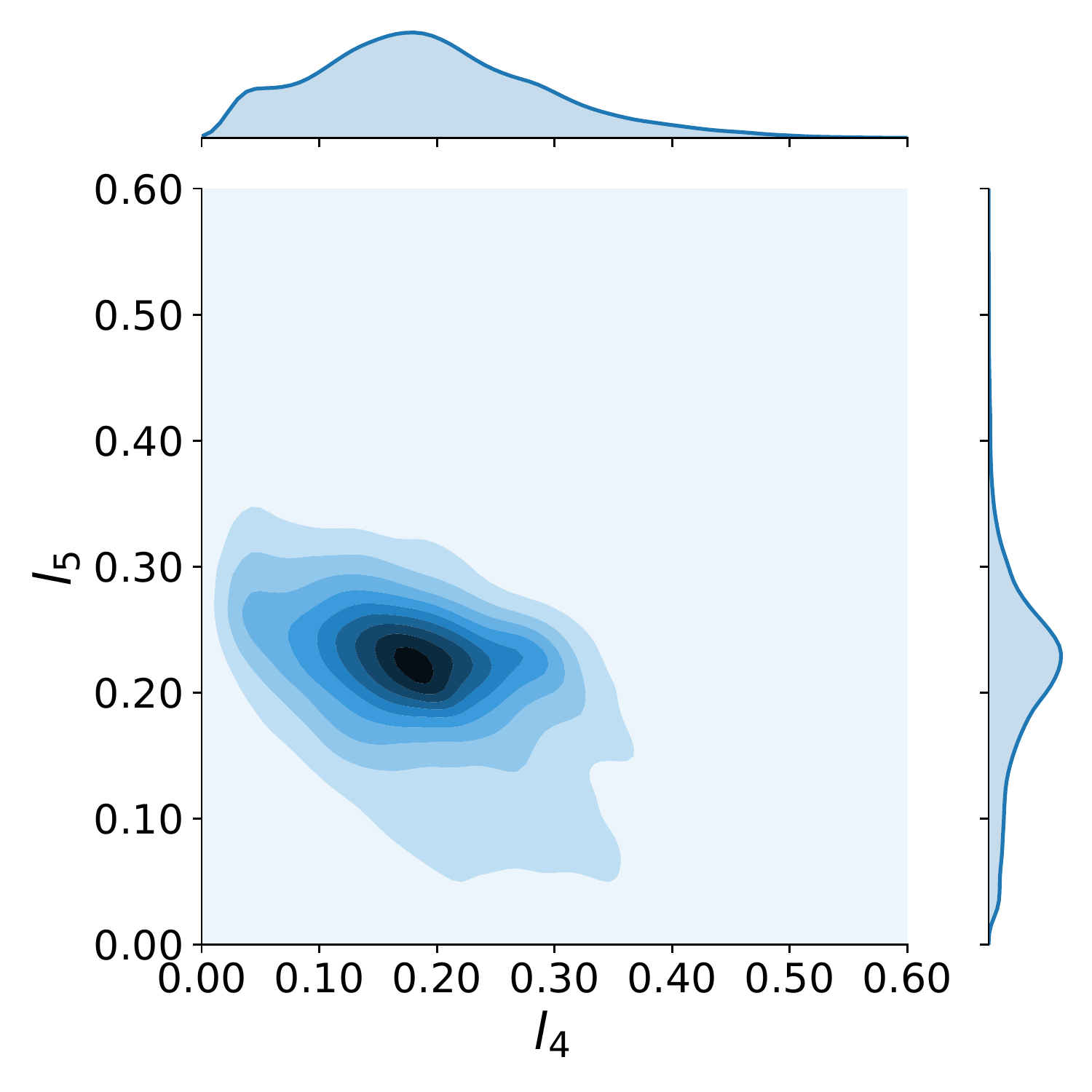}}
\caption{Joint density plots and corresponding marginal probabilities of $l_i$ and $l_{i+1}$ for profiles with five segments. $\rho$ is the Pearson correlation.}
\label{fig:jointl}
\end{figure*}

Figure~\ref{fig:jointl} shows the joint density plots for $l_i$ and $l_{i+1}$ and their respective Pearson correlations, $\rho$, which displayed moderate negative values. In general, the first segment is small, as illustrated in Figure~\ref{fig:jointl}(a), which can be an effect of the time of response, given by outside factors, such as dissemination on the web, conferences, advertisements on magazines, posts on social networks, media coverage, the popularity of the topic, and citations from other papers. Furthermore, similar outcomes were found for $\alpha_i$ and $\alpha_{i+1}$ (see Figure~\ref{fig:jointalpha} of supplementary material). In this case, we observe correlations that are not particularly strong. Two groups seem to be present on all the plots. A more detailed analysis of the groups in the profiles is performed in Section~\ref{sec:d}.
%\textcolor{red}{??? Nos dois casos aparecem dois grupos. Discutir junto.  A correlacao negativa indica que um segmento pequeno é seguido por um longo e vice-versa. Em geral inicia curta e em seguida fica longa e mais dispersa (ver histogramas e inclinacao). Pode indicar duracao da resposta e do tipo do fenomeno: (divulgacao na web, conferencia, anuncio em revista, redes sociais, cobertura na midia, popularidade do topico, citacoes.)}

\begin{figure*}[!htpb]
\centering
% \subfigure[Correlations between subsequent angles.]{\fbox{\includegraphics[scale=0.4]{imgs/alpha_alpha_corr.pdf}}}
% \subfigure[Correlations between subsequent segment extension pieces.]{\fbox{\includegraphics[scale=0.4]{imgs/l_l_corr.pdf}}}
% \subfigure[Correlations between angle and segment extension in the same or consecutive pieces.]
{{\includegraphics[scale=0.55]{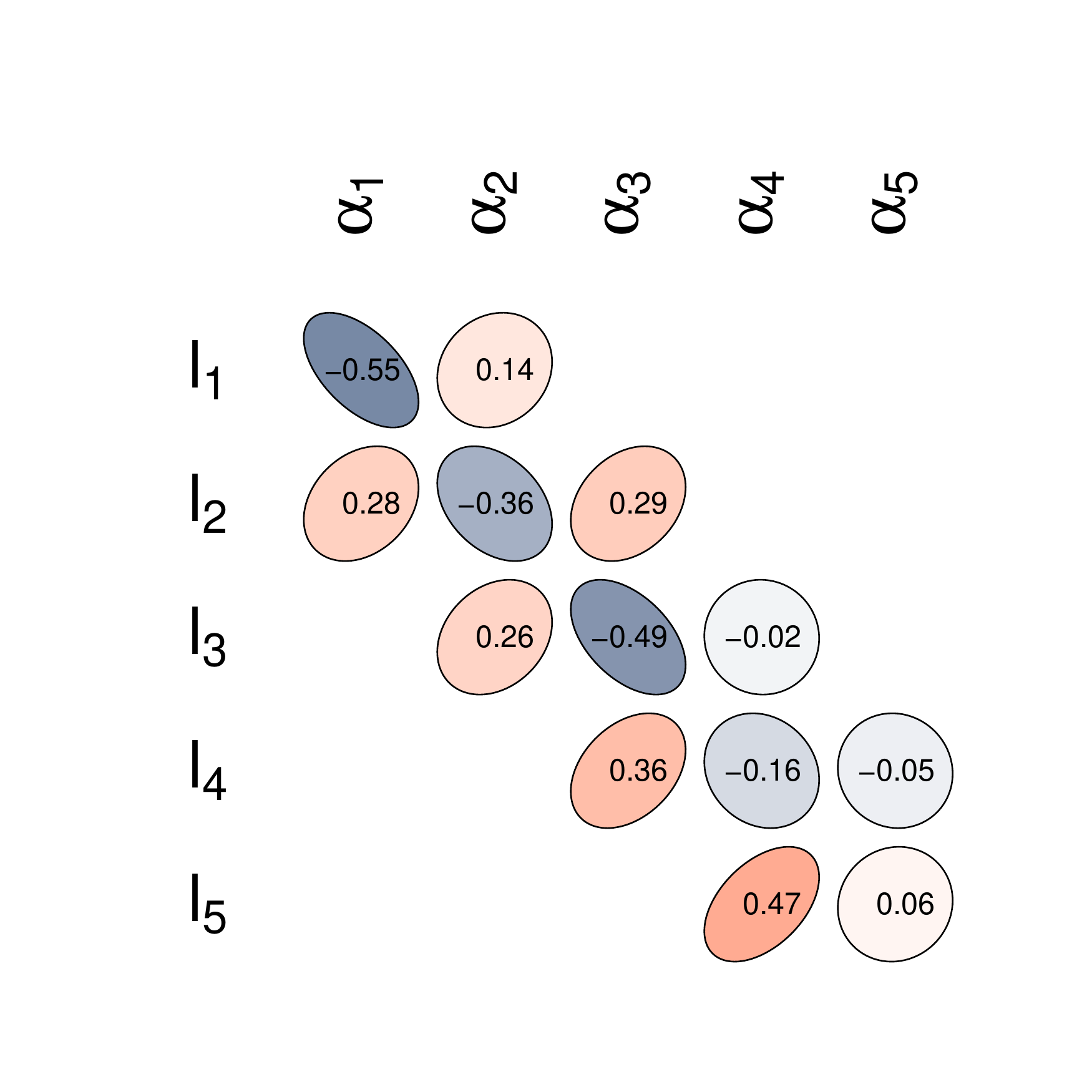}}}
\caption{Correlations between the variables of the piecewise curves shown as ellipses. The Pearson correlation coefficients were calculated by considering only the curves with five segments.
}
\label{fig:corr}
\end{figure*}

We also considered the relationships between $\alpha_i$ and $l_i$ at subsequent segments. The ellipses plots shown in Figure~\ref{fig:corr} represent the correlations between the different parameter combinations of subsequent segments. In more detail, the colors and inclination indicate the sign of the correlations, and negative correlations are plotted in blue while positive relations in red. The bigger the correlation magnitude, the stronger the color tone is, and more elongated are the ellipses. We found stronger correlations between $\alpha_i$ and $l_i$ in the initial segments compared to those found for the subsequent segments. In general, given a linear segment $i$, the correlations are negative: high inclination angles occur jointly with short segments, and lower inclination angles tend to occur for longer segments. Contrariwise, in consecutive segments, the correlation is positive. This result indicates that short segments tend to be followed by low inclination angles and long segments by high inclination angles. A possible explanation of this effect could be related to a sudden increase in views following the dissemination of the paper. In general, this surge of interest is not sustained along time.
%\textcolor{red}{Portanto, um evento abrupto momentaneo, divulgacao, anuncio. Em geral o primeiro momento gera subida rapida que não se sustenta. burst or surge of interest. Dificil manter momentos de burst por muito tempo. Depende da velocidade de difusao levando à saturacao da comunidade especifica. A midia pode ajudar, referencia na pagina da wiki e redes sociais.}

% (a) Pode não ser uma surpresa pois o método poligonal recupera o efeito zig-zag
% (c) Moderate correlation between lengths.
% Verificar na literatura aumentos de interesse após da primeira onda. (product life cycle). Tentar interpretar a estrutura poligonal como waves de life cycle.

%---------------------------------------------

\subsection{Clustering Analysis} \label{sec:d}
%Discutir que a modelagem poderia levar em conta a existencia de clusters com diferentes parametros.
% Discutir, em particular a identificacao de 2 principais grupos, mais um intermediario, que sao mais distintos para menos segmentos

Before proposing a stochastic model to reproduce the observed view profiles, it is interesting to check if we can find different patterns of view profiles that could be understood as clusters. This type of analysis can give us insights regarding the proposed model. More specifically, knowing about the existence of groups, it is possible to create separate models incorporating the singularities of the models. For that, a clustering analysis was performed in the measured segmented curves parameters. Groups were obtained by running a Hierarchical Clustering algorithm with the single-linkage criteria and considering Euclidean distances. We set the number of clusters to a maximum of three since other values led to less defined groups. For that analysis, each curve is represented by their set of segmented parameters $\alpha_i$ and $l_i$.

\begin{figure*}[!htpb]
    \centering
    \subfigure[][The 2-segment profiles clusters.]{\includegraphics[width=5cm]{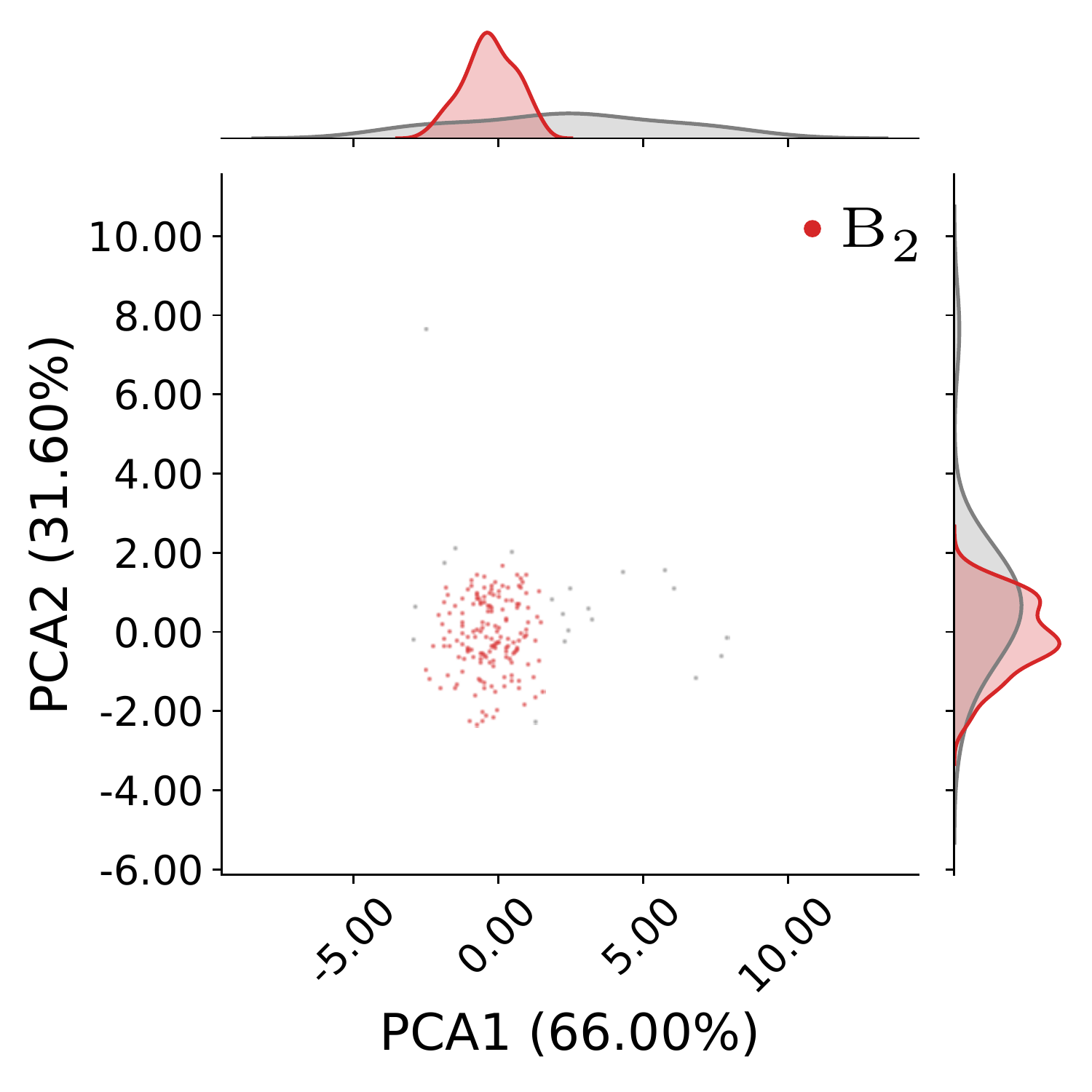}\label{fig:pca2}}
    \qquad
    \subfigure[][The 3-segment profiles clusters.]{\includegraphics[width=5cm]{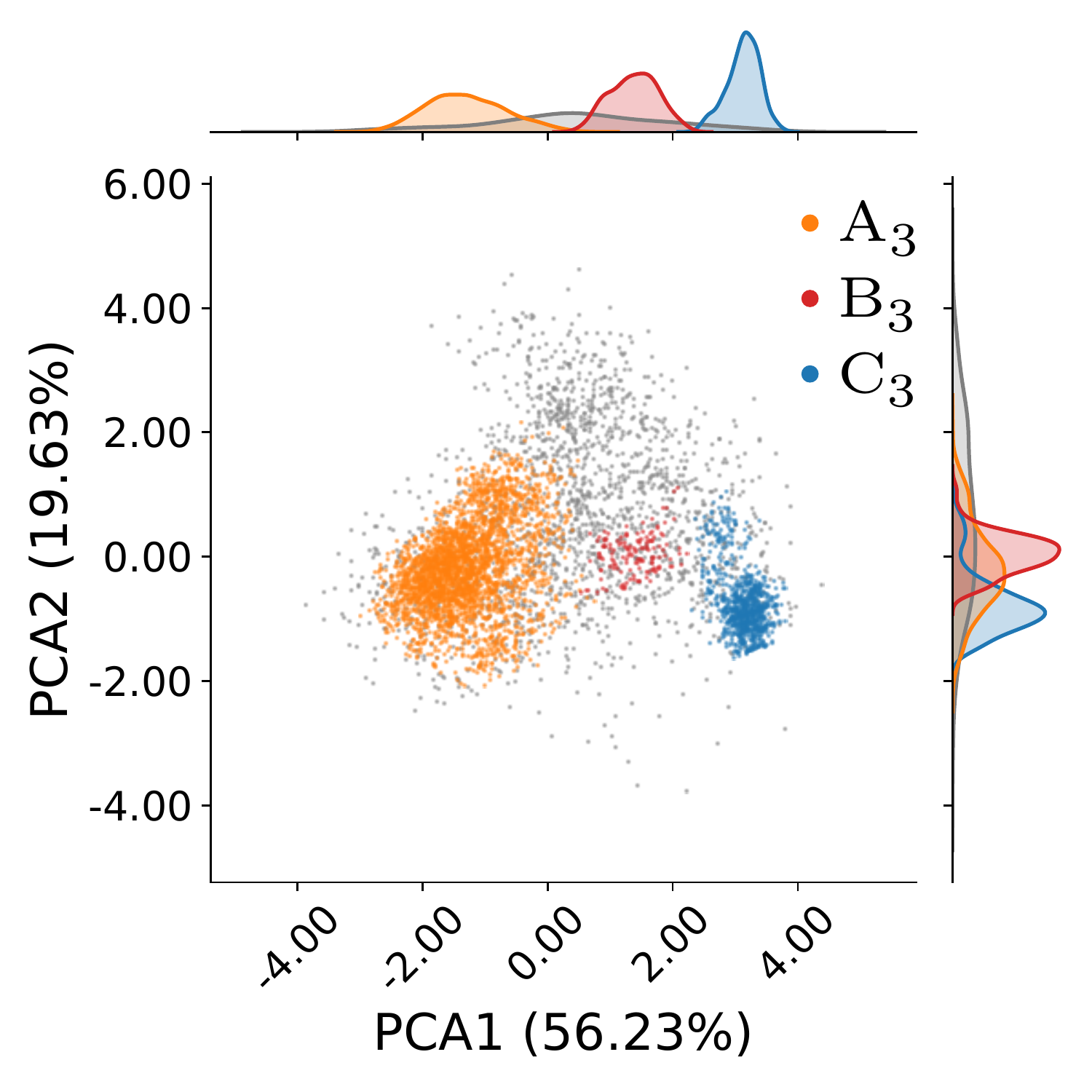}\label{fig:pca3}}
    \qquad
    \subfigure[][The 4-segment profiles clusters (PCA1 and PCA2).]{\includegraphics[width=5cm]{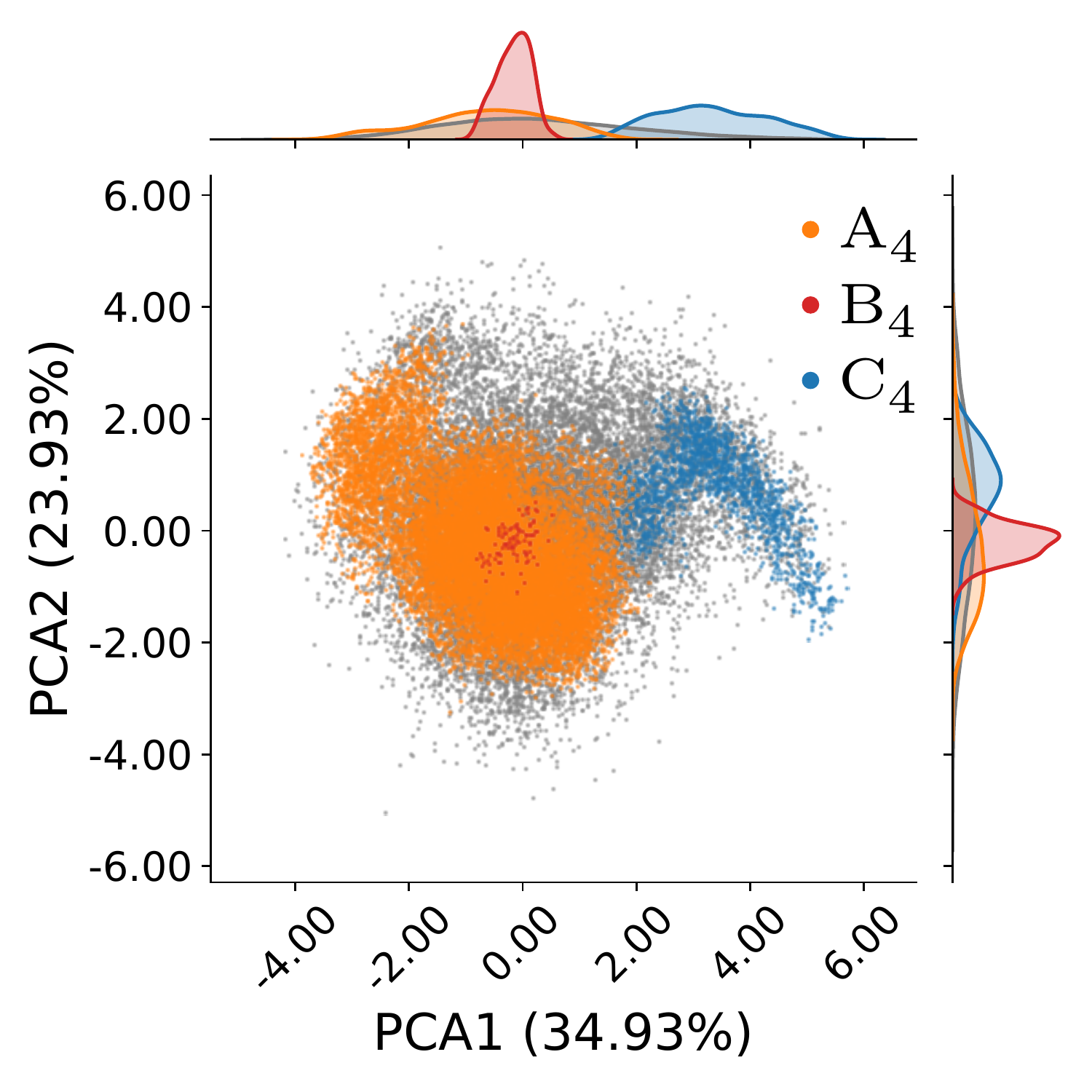}\label{fig:pca4c}}
    \qquad
    \subfigure[][The 4-segment profiles clusters (PCA1 and PCA3).]{\includegraphics[width=5cm]{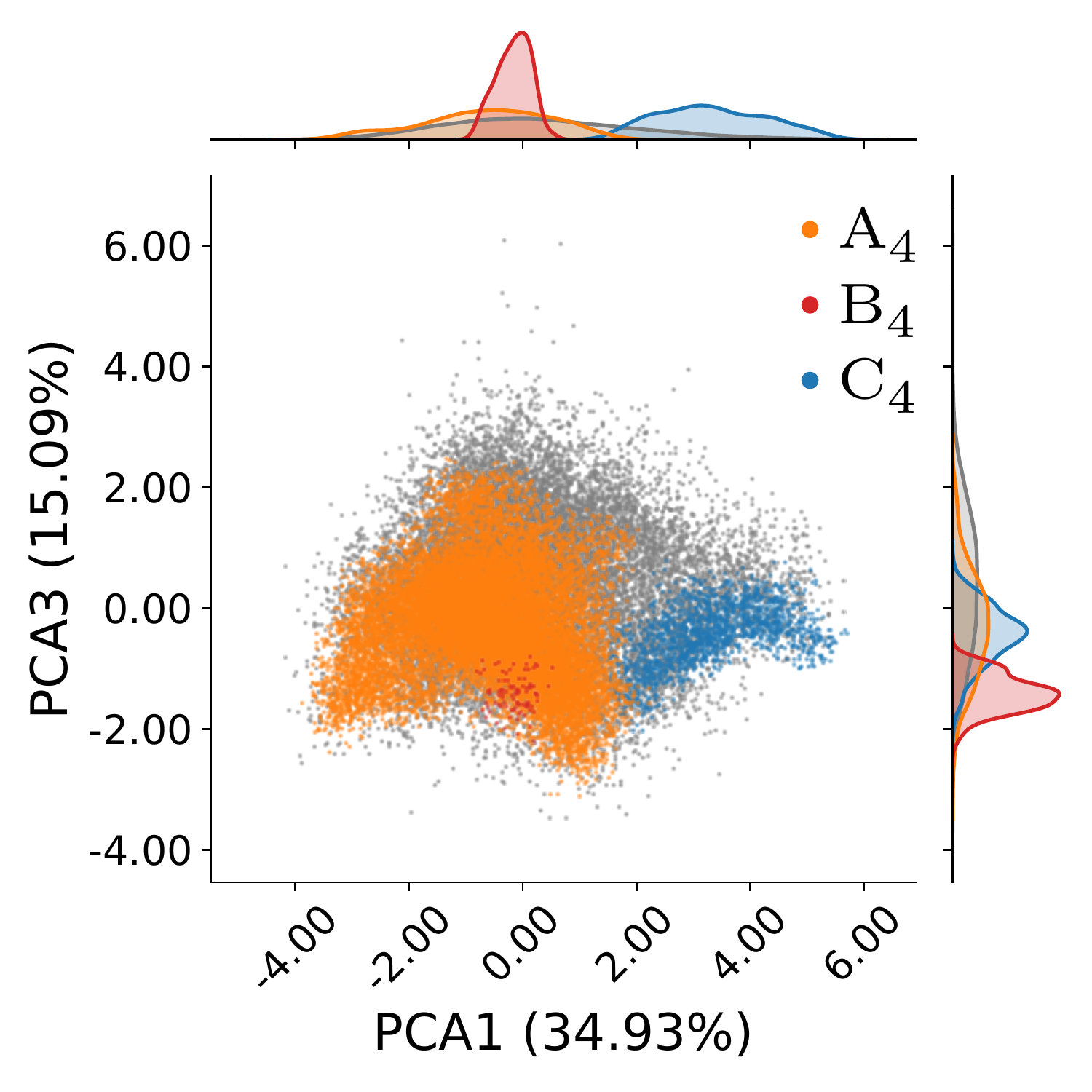}\label{fig:pca4d}}
    \qquad
    \subfigure[][The 5-segment profiles clusters (PCA1 and PCA2).]{\includegraphics[width=5cm]{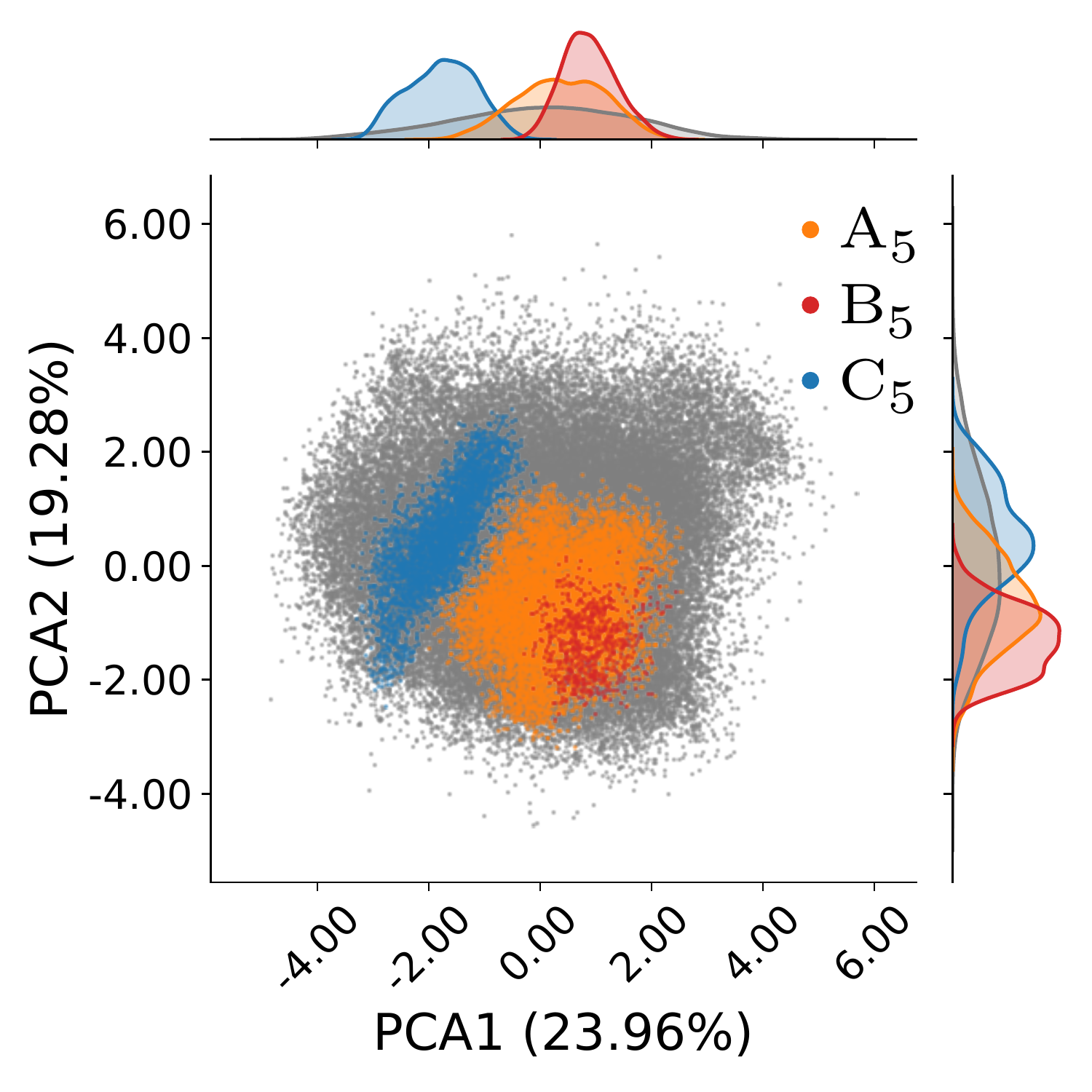}\label{fig:pca5e}}
    \qquad
    \subfigure[][The 5-segment profiles clusters (PCA1 and PCA3).]{\includegraphics[width=5cm]{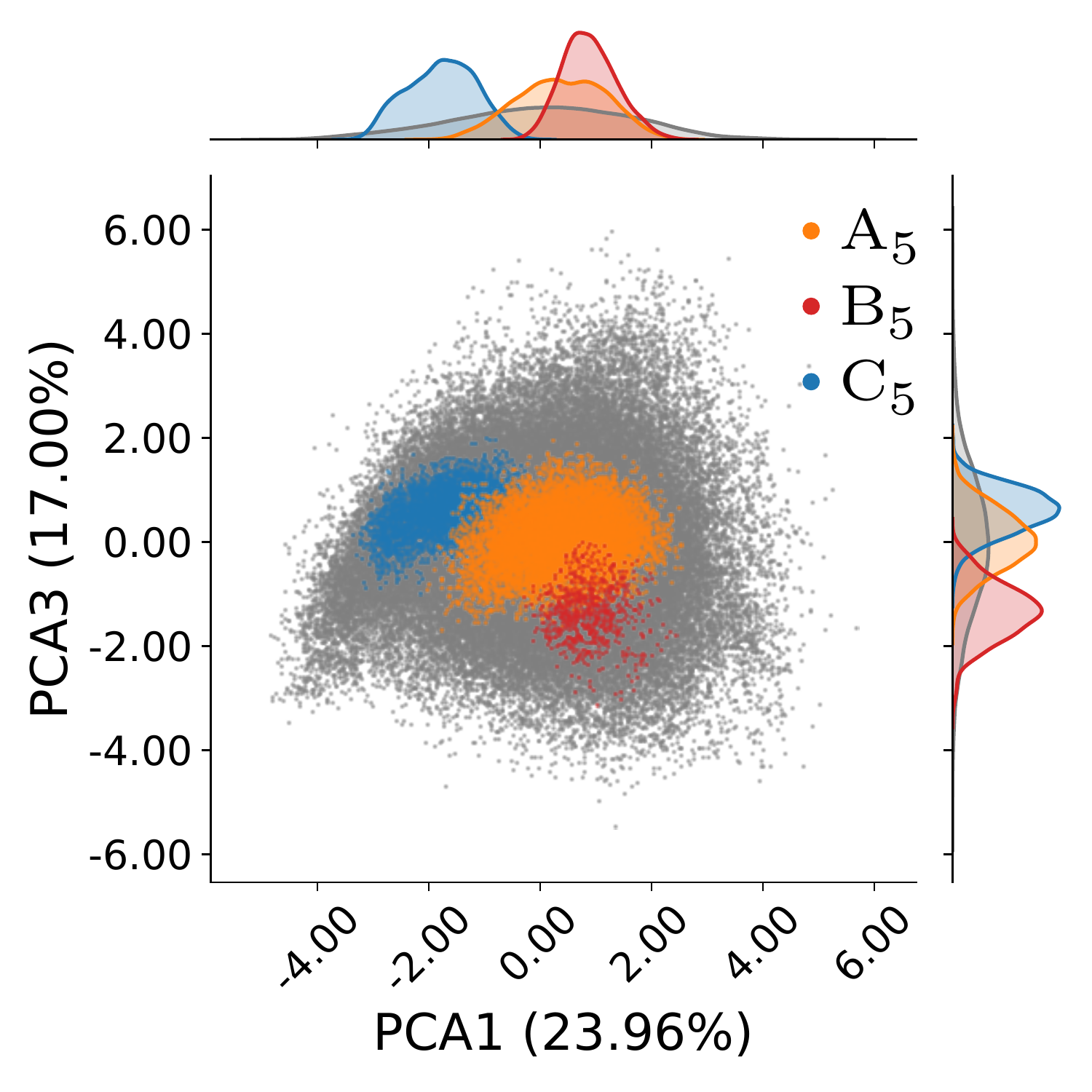}\label{fig:pca5f}}

    \caption{Scatter plots showing the clusters of the view profiles and the respective marginal distributions. The points correspond to the PCA projection of the angle and length of the segments of the profiles. The colors (orange, blue, and red) refer to the three detected groups, while the gray points correspond to views not assigned to these 3 groups. \label{fig:pcaclusters}}
\end{figure*}

The number of parameters defining the curves is at least four, corresponding to the simplest case in which only one breaking point exists. Thus, for view purposes, we employed a Principal Component Analysis (PCA) projection to reduce the dimensionality of the set of parameters~\citep{gewers}. The panels in Figure~\ref{fig:pcaclusters} show the clusters of curves for each of the adopted number of segments. In each of these scatterplots, each point represents a curve in the projected space. We named the three groups as $\text{A}_i$, $\text{B}_i$, and $\text{C}_i$, where $i$ refers to the number of segments. In the figure, the marginal distributions help to distinguish between the overlapping groups. Groups C tend to be more well-defined and separated from the others. The first two Principal Components accounts for 97.6\% of the total variance in Figure \ref{fig:pca2}, 75.86\% in Figure \ref{fig:pca3}, 58.86\% in Figures \ref{fig:pca4c} and \ref{fig:pca4d}, 43.24\% in Figures \ref{fig:pca5e} and \ref{fig:pca5f}. In the case of 2 and 3 segment clusters, only two principal components are enough to explain the variance of the data. However, for 4 and 5 segments, the third principal component is needed. This component is also shown in Figure~\ref{fig:pca2} for these cases.
%\textcolor{red}{Dizer que são altas e que 2 eixos explicam 2-3 segmentos enquanto 3 eixos explicam 4-5 segmentos.}

In general, the obtained clusters tend to overlap more as the number of segments increases. The marginal distributions are more separated when the curves have two or three segments in the PCA, their peaks being considerably distinct (Figures~\ref{fig:pca2}~and~\ref{fig:pca3}). When the curves have four or five segments, the distributions become flatter and indicate a larger overlap among the groups (Figures~\ref{fig:pca4c} and \ref{fig:pca4d}). One possible explanation for the change in the clustering structure is the progressive increase of view profile types with the number of segments.

\begin{figure*}
    \centering
    \subfigure[][Cluster $\text{A}_3$ (total of curves: 2677).]{\includegraphics[width=0.3\textwidth]{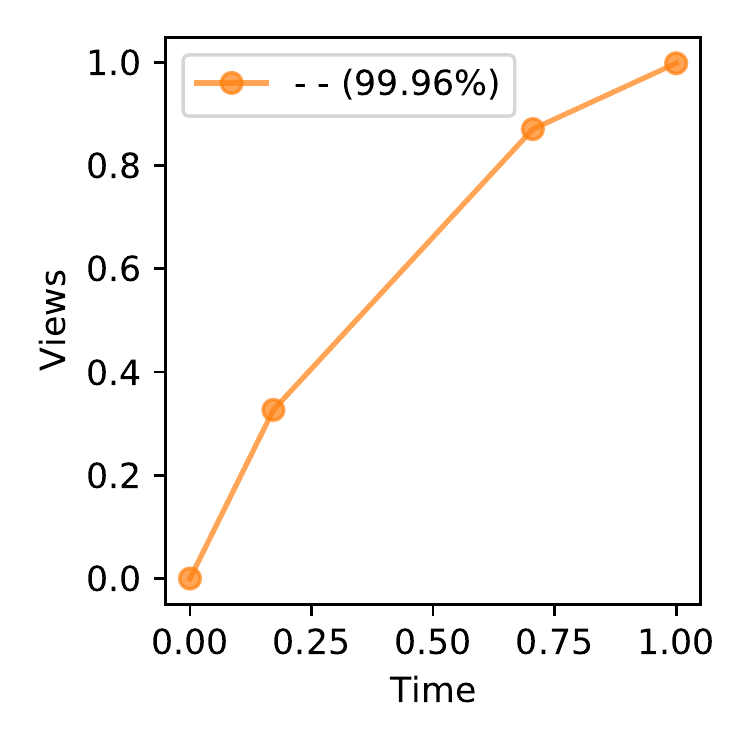}}
    \subfigure[][Cluster $\text{B}_3$ (total of curves: 143).]{\includegraphics[width=0.3\textwidth]{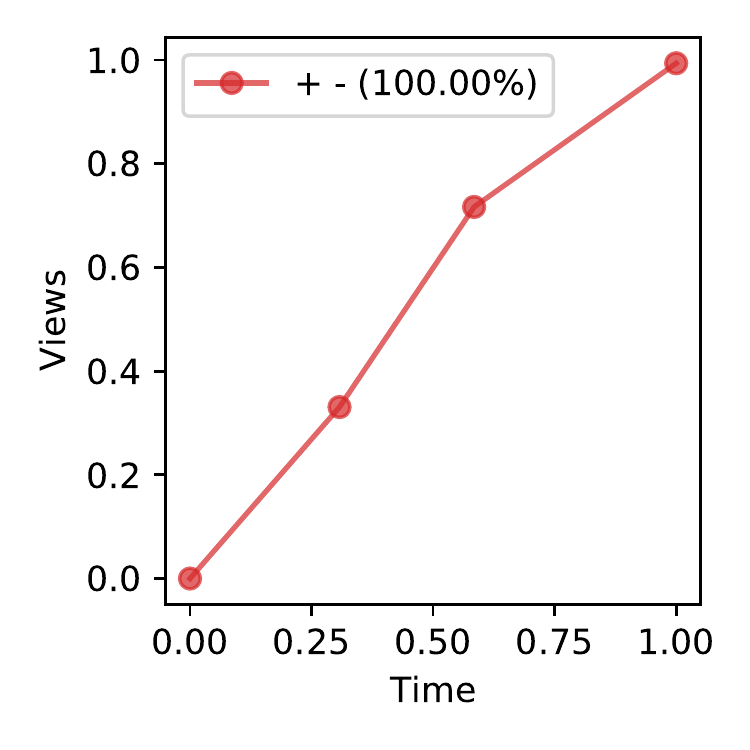}}
    \subfigure[][Cluster $\text{C}_3$ (total of curves: 804).]{\includegraphics[width=0.3\textwidth]{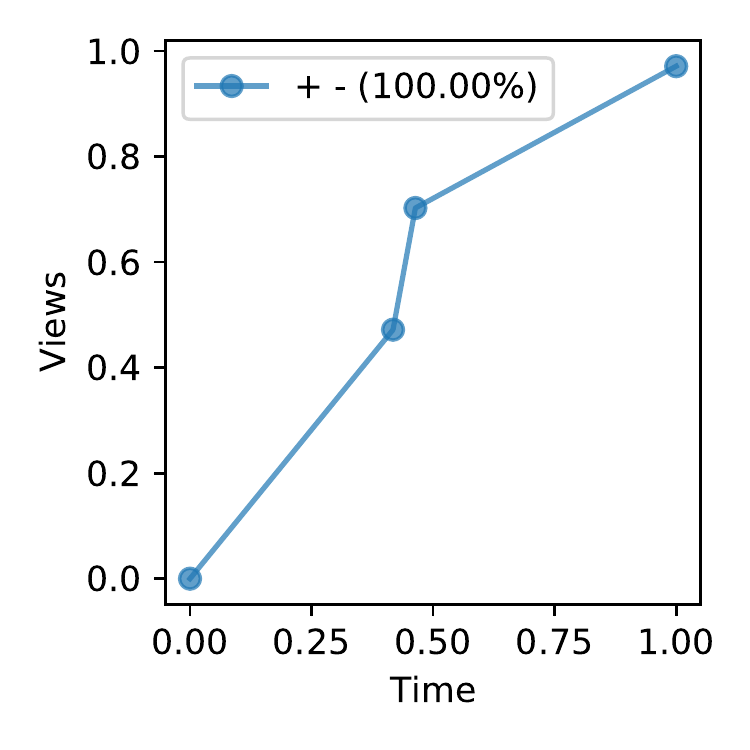}}

    \subfigure[][Cluster $\text{A}_5$ (total of curves: 7970).]{\includegraphics[width=0.3\textwidth]{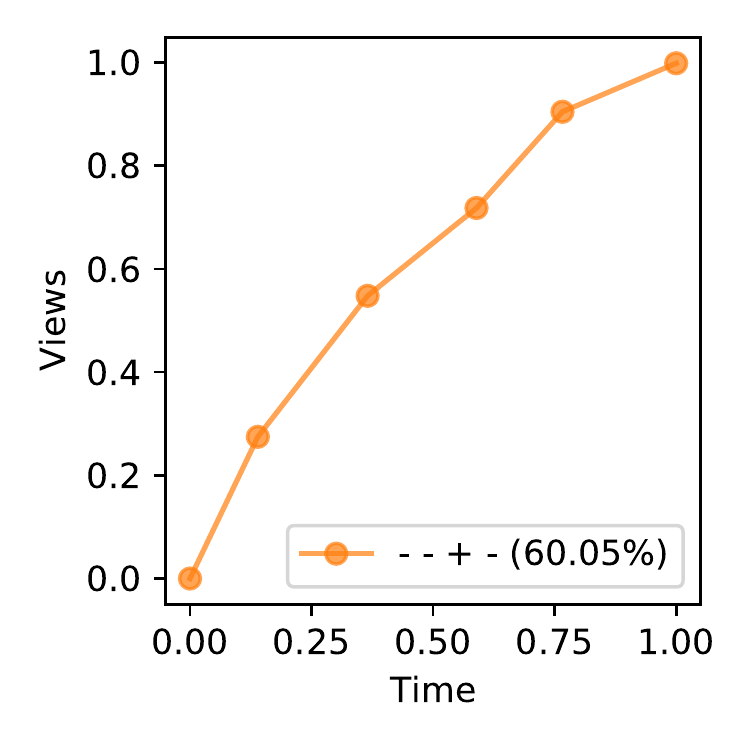}}
    \subfigure[][Cluster $\text{B}_5$ (total of curves: 661).]{\includegraphics[width=0.3\textwidth]{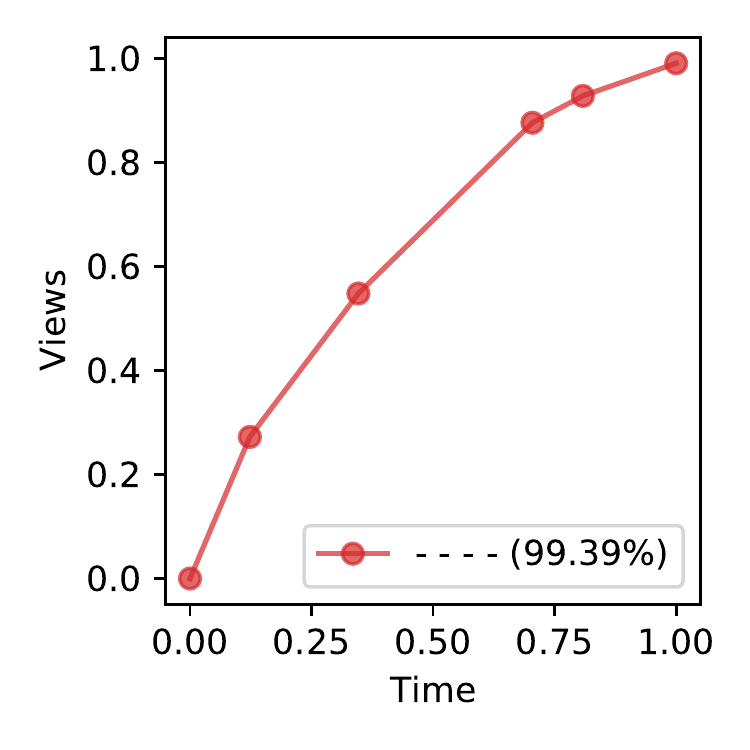}}
    \subfigure[][Cluster $\text{C}_5$ (total of curves: 3045).]{\includegraphics[width=0.3\textwidth]{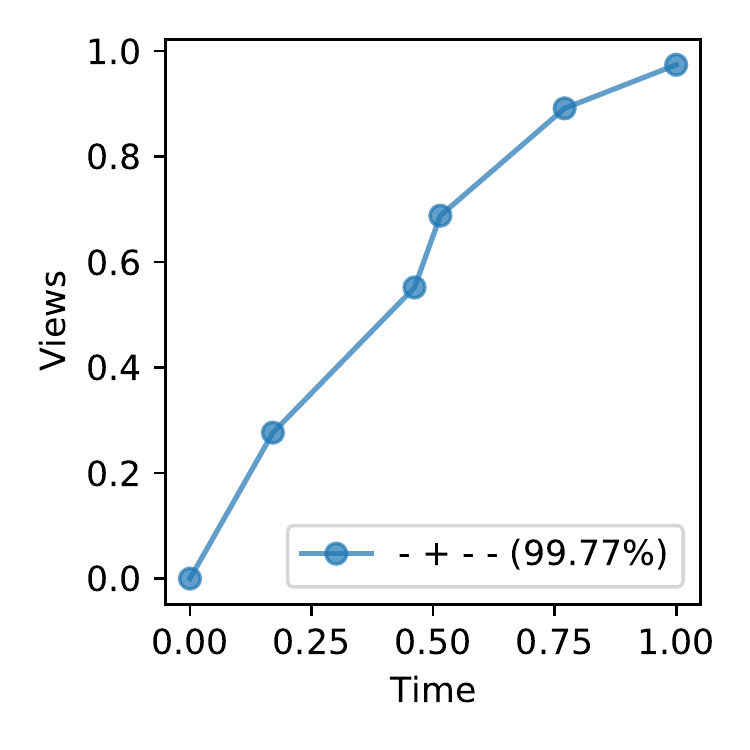}}

    \caption{Average curves obtained for the most frequent prototype of each cluster. The legend indicates the percentage of curves characterized by the most frequent prototype. The x-axis represents time, and the y-axis represents the cumulative number of views along time.}
    \label{fig:avecurves}
\end{figure*}

In the following analysis, we investigate and determine which are the typical features of the views of papers in groups $\{\text{A}_3,\text{A}_5\}$, $\{\text{B}_3,\text{B}_5\}$ and $\{\text{C}_3,\text{C}_5\}$. More specifically, the curve obtained for each view was assigned symbols ``$+$'' and ``$-$'' reflecting if the subsequent angle increased or decreased respectively to the previous angle of the previous segment.   Note that this analysis starts on the second angle. The resulting lists of ``$+$'' and ``$-$'' are then compared and organized into prototypes. For instance, in the case of 3 segments, we have: ``$--$'', ``$-+$'', ``$+-$'', and ``$++$''. The relative frequencies of each of these prototypes for papers in clusters A, B and C was then estimated, and the most frequent patterns are depicted in Figure~\ref{fig:avecurves}.  The signatures respective to the views with 3 and 5 segments are shown in Figures~\ref{fig:avecurves}(a-c) and (d-f), respectively.
In the case of the views with 3 segments, the most frequent group (a) is characterized by successive decreases of the view rates, indicating progressive assimilation by the respective research community.  Contrariwise, the situations in the groups shown in (b) and (c) are characterized by a relative increase of views along the intermediate segment.  This could have been implied by an event like dissemination in the news, meetings, or social networks. In the case of groups shown in (c), the increase in views occurs at a shorter time span than in (b).

The results for the articles with 5 segments obtained for the least frequent group (e), similarly to the group (a), was characterized by successive reductions of the viewing rate (``$----$'').  The other two groups, (d) and (f), are characterized by respective prototypes ``$--+-$'' and ``$-+--$''.  Therefore, cases (e) and (f) share the same prototype as (b) and (c).

% \begin{figure*}
%     \centering
%     \subfigure[][Average curves of the 3-segment profile clusters.]{\includegraphics[width=0.8\textwidth]{imgs/new_figs/ave_curve_3.pdf}}

%     \subfigure[][Average curves of the 5-segment profile clusters.]{\includegraphics[width=0.8\textwidth]{imgs/new_figs/ave_curve_5.pdf}}

%     \caption{Average curves obtained for the visualization profiles clusters. The x-axis represents time, and the y-axis represents the cumulative number of views along time.}
%     \label{fig:avecurves}
% \end{figure*}

% \begin{figure*}
%     \centering
%     \subfigure[][Average curves of the 3-segment profile clusters.]{\includegraphics[width=0.8\textwidth]{imgs/new_figs/ave_curve_3_intervals.pdf}}

%     \subfigure[][Average curves of the 5-segment profile clusters.]{\includegraphics[width=0.8\textwidth]{imgs/new_figs/ave_curve_5_intervals.pdf}}

%     \caption{INTERVALOS Average curves obtained for the visualization profiles clusters. The x-axis represents time, and the y-axis represents the cumulative number of views along time.}
%     \label{fig:avecurves}
% \end{figure*}

\subsection{Models Adherence} \label{sec:e}

\begin{figure*}[t!]
    \centering
    \includegraphics[width=0.95\linewidth]{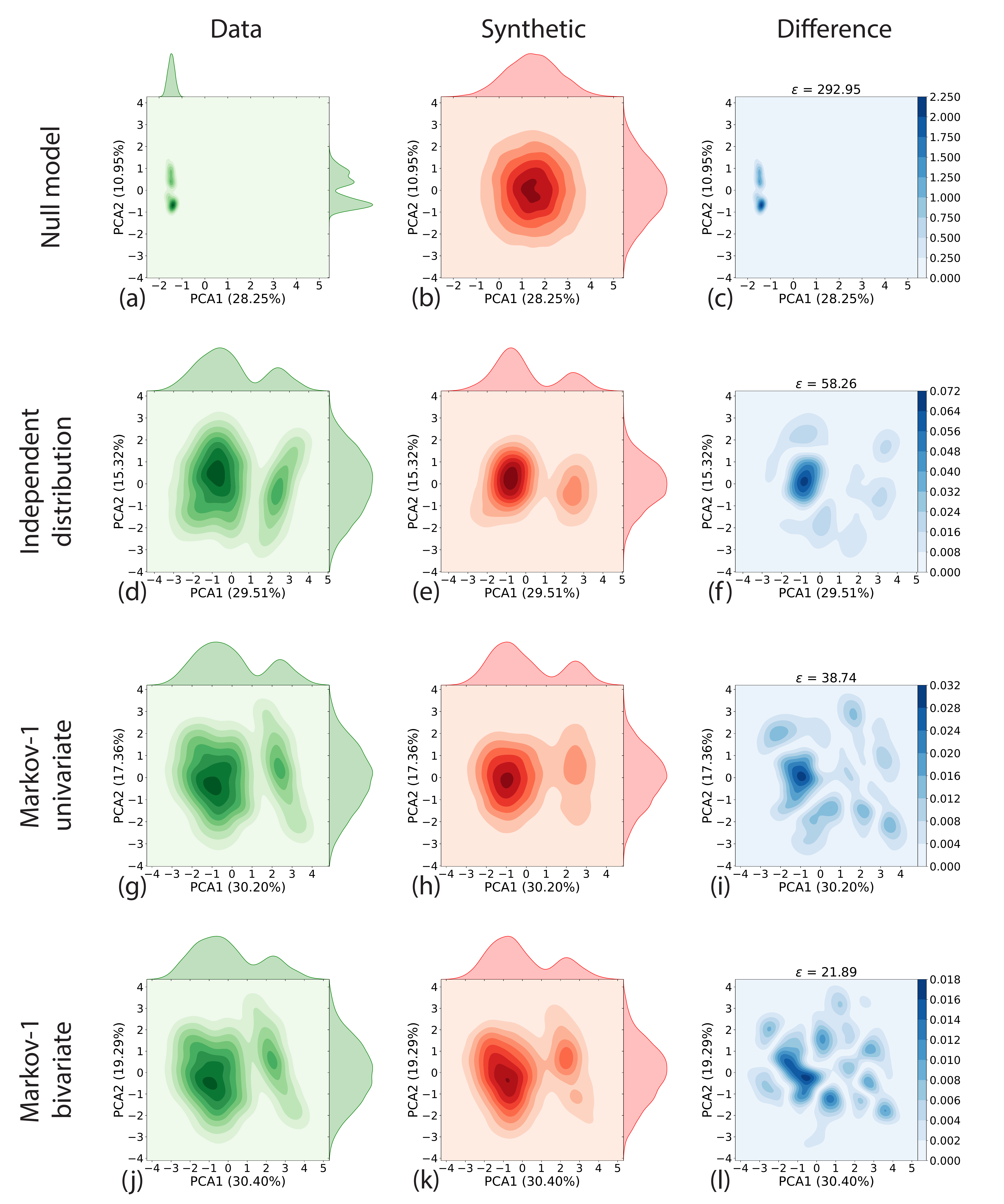}
    \caption{Comparison of the original profiles and the synthetic profiles produced according to the considered models. The three columns correspond respectively to the original data distributions along the PCA axes, the synthetic distributions and the difference. Note that the differences in density at the center of plot (c) are too small compared to the differences caused by the peaks of the original profiles. Note that for each model, a new PCA projection is obtained since it incorporate both the real and synthetic data.}  % (Scatter distance = 3.159945)
    \label{fig:models}
\end{figure*}

In this section, we check if the proposed Markov models can reproduce the joint distributions of the curves parameters ($\alpha_i$ and $l_i$) as expressed in terms of their principal component projections, shown in Figure~\ref{fig:models}. Observe that each line in this figure corresponds to one of the four considered models (see Section \ref{sec:statistical}). The marginal densities are also depicted along the respective axes.  For each of the four types of models, the cases corresponding to each of the three identified clusters were adjusted separately, and then combined when obtaining the principal component projection.

% Explicar a metodologia de validacao utilizando-se a diferenca das densidades das projecoes PCA

Then, the obtained density surfaces were compared by calculating absolute point-to-point differences ($\varepsilon$) between the original and synthetic 2d histograms of the PCA data (third column in Figure~\ref{fig:models}), and then adding all these values into the single error parameter $\varepsilon$, computed as
\begin{equation}
    \varepsilon = \sum_{x}\sum_{y} |\pi_{o}(x,y) - \pi_{s}(x,y)|,
    \label{eq:err} % ??? entre xmin e xmax 100 pontos, entre ymin e ymax 100 pontos
\end{equation}
where $\pi_{o}$ and $\pi_{s}$ are the surfaces corresponding to the original and synthetic data, respectively. Lower values of $\varepsilon$ indicate more accurate models.

As expected, the Null model resulted in the worst approximation of the curves (as seen in Figure~\ref{fig:models}(c)), and the quality of the models increases as we incorporate additional statistical information. The independent distribution model, shown in Figure~\ref{fig:models}(d-f), better approximates the original profiles without considering conditional probabilities (i.e.~memory). However, this model is unable to capture the medium scale details in the original distribution, and also broke the cluster. % os outros dois modelos foram ajustes finos que proporcionaram dados sinteticos mais parecidos com os dados reais OK

The models based on Markov-1 take into account not only the independent conditional distributions but also the parameters of consecutive segments. The Markov-1 univariate synthetic profiles resulted in better approximations of the real view profiles (Figure~\ref{fig:models}(g-i)).
Finally, the Markov-1 multivariate model produced synthetic profiles more similar to the original counterparts (Figure~\ref{fig:models}(j-l)). The bivariate distribution of $\alpha$ and $l$ provide the best approximation of the original data among the considered models, as it yields the smallest $\varepsilon$ value. This suggests that not only the angles of previous segments but also their lengths are important subsidies in predicting the parameters of the next segments.

All in all, the obtained results confirm to a good extent the initial hypothesis that the view profiles are not trivial or random and have intrinsic structure that were progressively reflected by modeling approaches taking into account additional information.   This implies that there are interesting real-world effects and mechanisms implementing the types of observed structure.  In particular, we have memory effects and time dependencies, in the sense that the properties of one segment tend to correlated with subsequent segments.  These effects could be hypothetically related to tendencies such as a brief surge of interest, e.g.~caused by media dissemination, would be followed by a longer period of less intense inclination.

\section{Concluding Remarks}
\label{sec: concluding}

Science is inherently a collaborative endeavor. Therefore, the speed of dissemination of new ideas has a relevant impact in the development of novel theories and experiments. Traditionally, the impact of papers has been studied in terms of the number of citations, but since the World Wide Web became the main medium for publishing papers, the development of new data aggregation tools led to the definition of many alternative metrics~\citep{sud2014evaluating}. One of the simplest of such metrics is the number of page views. Measuring page views is relatively simple and can usually be done with arbitrary granularity (hourly, daily, monthly, etc). Compared to citations, the number of views also tends to display a much lower delay to important events such as publication and conference presentation.

Here, we studied to what extent the monthly number of views for articles published in the PLoS ONE journal present a polygonal structure, which constitute the main question of the present work. A key observation regarding the number of views of the articles in the PLoS ONE dataset was used in the analysis: articles tend to display periods of relatively constant number of monthly views, with sharp changes in views between such periods. This hypothesis was investigated throughout the work by considering the cumulative number of article views. If the hypothesis is true, the cumulative number of views should be correctly represented by a piece-wise linear function.

A segmented least squares regression methodology~\citep{muggeo2003estimating} was applied to identify breakpoints between linear segments in cumulative article views, and the length $l_i$ and angle $\alpha_i$ of each segment were measured and used as parameters of four models for generating synthetic article views profiles. The models took into account progressively more information about the profiles, so as to allow the identification of the most relevant properties.

Several interesting results were obtained. It was verified that the segmented regression led to a lower RMSE than in the case of synthetic profiles generated from a model which took into account randomly generated number of monthly views. The result indicates that representing the cumulative number of article views by a piece-wise linear function led to a relatively low regression error. Thus, the profiles can be modeled by linear segments. Another important result was the observation of two view profiles, one corresponding to a relatively low initial slope followed by a higher slope and another presenting only slopes that decrease with time. In order to better interpret these groups, additional metadata is necessary and is a future development of this work. Regarding the synthetic models, it was found that curves generated from $\alpha_i$ and $l_i$ sampled independently with the same distribution as the real data led to profiles that approximated well the real profiles. Taking into account conditional probabilities between subsequent segments and between $\alpha_i$ and $l_i$ led to improved models. Although we found many interesting results, our study also includes some limitations. The analysis was performed only with a single dataset, obtained from PLoS ONE papers, and cannot be generalized for all journals.

For future developments, additional metadata about the papers can be taken into account in the analysis. In particular, it would be interesting to investigate how the views dynamics changes according to the subject area and authors institution. It would also be interesting to associate social network data with the observed profiles. For instance, verify if the identified breakpoints correlate with messages published by power users~\citep{eysenbach2011can} in a social network about the article. It is also worth investigating how bibliometric networks can affect view patterns along time~\citep{de2017knowledge,da2006learning}.
Finally, we could also analyze if other factors such as authors and topics  visibility can affect the patterns of view profiles in papers~\citep{correa2017patterns,lariviere2016contributorship,amancio2015comparing,lu2019analyzing}.

%Resultados:
%- Estatisticas basicas: quantos papers, tempo em anos.....

%- Fiting das curvas: serve pra uma certa quantidade de papers, exemplos de onde isso acontece e mostrar onde ta quebrando

%- Distribuicao do erro, e falar eliminou a partir daquele erro (ou seja, nao sao todos que sao como retas)

%-> Estatisticas de quebra: quantidade maxima de quebra foi 5 e o metodo escolhe a melhor.

%-> Mostrar correlacoes angulo angulo , comprimento angulo

%-> Mostrar densidades, Cesar: explica isso pro Filipi (angulos sao afetados)

%-> Filipi: fala sobre o modelo
%- PCA usando modelo aleatorio
%- PCA usando o modelo que ja tem
%- PCA usando: "sortear da distribuicao independente" (memoria zero)

\newpage

\section*{Acknowledgements}

This study was financed in part by the Coordena\c {c}\~{a}o de Aperfei\c{c}oamento de Pessoal de N\'{i}vel Superior -- Brasil (CAPES) -- Finance Code 001.
C. H. Comin thanks FAPESP (grant no. 18/09125-4) for financial support. F. N. Silva acknowledges CAPES and FAPESP (grant no. 15/08003-4). H. F. de Arruda acknowledges FAPESP for sponsorship (grants 2018/10489-0 and 2019/16223-5). D. R. Amancio thanks FAPESP (grant no. 16/19069-9) and CNPq (grant no.  304026/2018-2).
L. da F. Costa thanks CNPq (grant no.  307085/2018-0) and NAP-PRP-USP for support. This work has been supported also by the FAPESP grant 15/22308-2.

\newpage

\bibliographystyle{abbrvnat}

%\bibliography{references}

\begin{thebibliography}{43}
\providecommand{\natexlab}[1]{#1}
\providecommand{\url}[1]{\texttt{#1}}
\expandafter\ifx\csname urlstyle\endcsname\relax
  \providecommand{\doi}[1]{doi: #1}\else
  \providecommand{\doi}{doi: \begingroup \urlstyle{rm}\Url}\fi

\bibitem[Amancio(2015)]{amancio2015comparing}
D.~R. Amancio.
\newblock Comparing the topological properties of real and artificially
  generated scientific manuscripts.
\newblock \emph{Scientometrics}, 105\penalty0 (3):\penalty0 1763--1779, 2015.

\bibitem[Amancio et~al.(2012)Amancio, Oliveira~Jr, and Costa]{2012three}
D.~R. Amancio, O.~N. Oliveira~Jr, and L.~F. Costa.
\newblock Three-feature model to reproduce the topology of citation networks
  and the effects from authors' visibility on their h-index.
\newblock \emph{Journal of Informetrics}, 6\penalty0 (3):\penalty0 427--434,
  2012.

\bibitem[Bollen et~al.(2005)Bollen, Van~de Sompel, Smith, and
  Luce]{bollen2005toward}
J.~Bollen, H.~Van~de Sompel, J.~A. Smith, and R.~Luce.
\newblock Toward alternative metrics of journal impact: A comparison of
  download and citation data.
\newblock \emph{Information processing \& management}, 41\penalty0
  (6):\penalty0 1419--1440, 2005.

\bibitem[Bornmann(2014)]{bornmann2014altmetrics}
L.~Bornmann.
\newblock Do altmetrics point to the broader impact of research? an overview of
  benefits and disadvantages of altmetrics.
\newblock \emph{Journal of Informetrics}, 8\penalty0 (4):\penalty0 895--903,
  2014.

\bibitem[Brody et~al.(2006)Brody, Harnad, and Carr]{brody2006earlier}
T.~Brody, S.~Harnad, and L.~Carr.
\newblock Earlier web usage statistics as predictors of later citation impact.
\newblock \emph{Journal of the American Society for Information Science and
  Technology}, 57\penalty0 (8):\penalty0 1060--1072, 2006.

\bibitem[Chen et~al.(2020)Chen, Deng, Zhong, and Zhang]{chen2020exploring}
B.~Chen, D.~Deng, Z.~Zhong, and C.~Zhang.
\newblock Exploring linguistic characteristics of highly browsed and downloaded
  academic articles.
\newblock \emph{Scientometrics}, 122\penalty0 (3):\penalty0 1769--1790, 2020.

\bibitem[Corr{\^e}a~Jr et~al.(2017)Corr{\^e}a~Jr, Silva, Costa, and
  Amancio]{correa2017patterns}
E.~A. Corr{\^e}a~Jr, F.~N. Silva, L.~F. Costa, and D.~R. Amancio.
\newblock Patterns of authors contribution in scientific manuscripts.
\newblock \emph{Journal of Informetrics}, 11\penalty0 (2):\penalty0 498--510,
  2017.

\bibitem[Costa(2006)]{da2006learning}
L.~F. Costa.
\newblock Learning about knowledge: A complex network approach.
\newblock \emph{Physical Review E}, 74\penalty0 (2):\penalty0 026103, 2006.

\bibitem[de~Arruda et~al.(2017)de~Arruda, Silva, Costa, and
  Amancio]{de2017knowledge}
H.~F. de~Arruda, F.~N. Silva, L.~F. Costa, and D.~R. Amancio.
\newblock Knowledge acquisition: A complex networks approach.
\newblock \emph{Information Sciences}, 421:\penalty0 154--166, 2017.

\bibitem[de~Winter(2015)]{de2015relationship}
J.~C. de~Winter.
\newblock The relationship between tweets, citations, and article views for
  plos one articles.
\newblock \emph{Scientometrics}, 102\penalty0 (2):\penalty0 1773--1779, 2015.

\bibitem[Duan and Xiong(2017)]{duan2017download}
Y.~Duan and Z.~Xiong.
\newblock Download patterns of journal papers and their influencing factors.
\newblock \emph{Scientometrics}, 112\penalty0 (3):\penalty0 1761--1775, 2017.

\bibitem[Efron et~al.(2004)Efron, Hastie, Johnstone, Tibshirani,
  et~al.]{efron2004least}
B.~Efron, T.~Hastie, I.~Johnstone, R.~Tibshirani, et~al.
\newblock Least angle regression.
\newblock \emph{The Annals of Statistics}, 32\penalty0 (2):\penalty0 407--499,
  2004.

\bibitem[Erdt et~al.(2016)Erdt, Nagarajan, Sin, and Theng]{erdt2016altmetrics}
M.~Erdt, A.~Nagarajan, S.-C.~J. Sin, and Y.-L. Theng.
\newblock Altmetrics: an analysis of the state-of-the-art in measuring research
  impact on social media.
\newblock \emph{Scientometrics}, 109\penalty0 (2):\penalty0 1117--1166, 2016.

\bibitem[Eysenbach(2011)]{eysenbach2011can}
G.~Eysenbach.
\newblock Can tweets predict citations? metrics of social impact based on
  twitter and correlation with traditional metrics of scientific impact.
\newblock \emph{Journal of medical Internet research}, 13\penalty0
  (4):\penalty0 e123, 2011.

\bibitem[Galligan and Dyas-Correia(2013)]{galligan2013altmetrics}
F.~Galligan and S.~Dyas-Correia.
\newblock Altmetrics: Rethinking the way we measure.
\newblock \emph{Serials review}, 39\penalty0 (1):\penalty0 56--61, 2013.

\bibitem[Gewers et~al.(2018)Gewers, Ferreira, de~Arruda, Silva, Comin, Amancio,
  and Costa]{gewers}
F.~L. Gewers, G.~R. Ferreira, H.~F. de~Arruda, F.~N. Silva, C.~H. Comin, D.~R.
  Amancio, and L.~F. Costa.
\newblock Principal component analysis: A natural approach to data exploration.
\newblock \emph{arXiv}, 1804.02502\penalty0 (1):\penalty0 1--33, 2018.

\bibitem[Huang et~al.(2018)Huang, Wang, and Wu]{huang2018altmetric}
W.~Huang, P.~Wang, and Q.~Wu.
\newblock A correlation comparison between altmetric attention scores and
  citations for six plos journals.
\newblock \emph{PLOS ONE}, 13\penalty0 (4):\penalty0 1--15, 04 2018.
\newblock \doi{10.1371/journal.pone.0194962}.
\newblock URL \url{https://doi.org/10.1371/journal.pone.0194962}.

\bibitem[Hyndman and Koehler(2006)]{hyndman2006another}
R.~J. Hyndman and A.~B. Koehler.
\newblock Another look at measures of forecast accuracy.
\newblock \emph{International journal of forecasting}, 22\penalty0
  (4):\penalty0 679--688, 2006.

\bibitem[Ioannidis et~al.(2014)Ioannidis, Boyack, Small, Sorensen, and
  Klavans]{ioannidis2014bibliometrics}
J.~Ioannidis, K.~W. Boyack, H.~Small, A.~A. Sorensen, and R.~Klavans.
\newblock Bibliometrics: Is your most cited work your best?
\newblock \emph{Nature News}, 514\penalty0 (7524):\penalty0 561, 2014.

\bibitem[Jain et~al.(1999)Jain, Murty, and Flynn]{jain1999data}
A.~K. Jain, M.~N. Murty, and P.~J. Flynn.
\newblock Data clustering: a review.
\newblock \emph{ACM computing surveys (CSUR)}, 31\penalty0 (3):\penalty0
  264--323, 1999.

\bibitem[Jamali and Nikzad(2011)]{jamali2011article}
H.~R. Jamali and M.~Nikzad.
\newblock Article title type and its relation with the number of downloads and
  citations.
\newblock \emph{Scientometrics}, 88\penalty0 (2):\penalty0 653--661, 2011.

\bibitem[Larivi{\`e}re et~al.(2013)Larivi{\`e}re, Ni, Gingras, Cronin, and
  Sugimoto]{lariviere2013bibliometrics}
V.~Larivi{\`e}re, C.~Ni, Y.~Gingras, B.~Cronin, and C.~R. Sugimoto.
\newblock Bibliometrics: Global gender disparities in science.
\newblock \emph{Nature News}, 504\penalty0 (7479):\penalty0 211, 2013.

\bibitem[Larivi{\`e}re et~al.(2016)Larivi{\`e}re, Desrochers, Macaluso,
  Mongeon, Paul-Hus, and Sugimoto]{lariviere2016contributorship}
V.~Larivi{\`e}re, N.~Desrochers, B.~Macaluso, P.~Mongeon, A.~Paul-Hus, and
  C.~R. Sugimoto.
\newblock Contributorship and division of labor in knowledge production.
\newblock \emph{Social Studies of Science}, 46\penalty0 (3):\penalty0 417--435,
  2016.

\bibitem[Lu et~al.(2019)Lu, Bu, Dong, Wang, Ding, Larivi{\`e}re, Sugimoto,
  Paul, and Zhang]{lu2019analyzing}
C.~Lu, Y.~Bu, X.~Dong, J.~Wang, Y.~Ding, V.~Larivi{\`e}re, C.~R. Sugimoto,
  L.~Paul, and C.~Zhang.
\newblock Analyzing linguistic complexity and scientific impact.
\newblock \emph{Journal of Informetrics}, 13\penalty0 (3):\penalty0 817--829,
  2019.

\bibitem[Moed(2005)]{moed2005statistical}
H.~F. Moed.
\newblock Statistical relationships between downloads and citations at the
  level of individual documents within a single journal.
\newblock \emph{Journal of the American Society for Information Science and
  Technology}, 56\penalty0 (10):\penalty0 1088--1097, 2005.

\bibitem[Muggeo(2003)]{muggeo2003estimating}
V.~M. Muggeo.
\newblock Estimating regression models with unknown break-points.
\newblock \emph{Statistics in medicine}, 22\penalty0 (19):\penalty0 3055--3071,
  2003.

\bibitem[Muggeo and Adelfio(2010)]{muggeo2010efficient}
V.~M. Muggeo and G.~Adelfio.
\newblock Efficient change point detection for genomic sequences of continuous
  measurements.
\newblock \emph{Bioinformatics}, 27\penalty0 (2):\penalty0 161--166, 2010.

\bibitem[M{\"u}llner(2011)]{mullner2011modern}
D.~M{\"u}llner.
\newblock Modern hierarchical, agglomerative clustering algorithms.
\newblock \emph{arXiv preprint arXiv:1109.2378}, 2011.

\bibitem[Note1()]{Note1}
Note1.
\newblock \protect \url {https://journals.plos.org/plosone}.

\bibitem[Ortega(2018)]{ortega2018life}
J.~L. Ortega.
\newblock The life cycle of altmetric impact: A longitudinal study of six
  metrics from plumx.
\newblock \emph{Journal of Informetrics}, 12\penalty0 (3):\penalty0 579--589,
  2018.

\bibitem[Peat(2002)]{peat2002certainty}
F.~D. Peat.
\newblock \emph{From certainty to uncertainty: The story of science and ideas
  in the twentieth century}.
\newblock Joseph Henry Press, 2002.

\bibitem[Perneger(2004)]{perneger2004relation}
T.~V. Perneger.
\newblock Relation between online “hit counts” and subsequent citations:
  prospective study of research papers in the bmj.
\newblock \emph{Bmj}, 329\penalty0 (7465):\penalty0 546--547, 2004.

\bibitem[Priem et~al.(2012)Priem, Piwowar, and Hemminger]{priem2012altmetrics}
J.~Priem, H.~A. Piwowar, and B.~M. Hemminger.
\newblock Altmetrics in the wild: Using social media to explore scholarly
  impact.
\newblock \emph{arXiv preprint arXiv:1203.4745}, 2012.

\bibitem[Schl{\"o}gl et~al.(2014)Schl{\"o}gl, Gorraiz, Gumpenberger, Jack, and
  Kraker]{schlogl2014comparison}
C.~Schl{\"o}gl, J.~Gorraiz, C.~Gumpenberger, K.~Jack, and P.~Kraker.
\newblock Comparison of downloads, citations and readership data for two
  information systems journals.
\newblock \emph{Scientometrics}, 101\penalty0 (2):\penalty0 1113--1128, 2014.

\bibitem[Shuai et~al.(2012)Shuai, Pepe, and Bollen]{shuai2012scientific}
X.~Shuai, A.~Pepe, and J.~Bollen.
\newblock How the scientific community reacts to newly submitted preprints:
  Article downloads, twitter mentions, and citations.
\newblock \emph{PloS one}, 7\penalty0 (11), 2012.

\bibitem[Sokal and Rohlf(1962)]{sokal1962comparison}
R.~R. Sokal and F.~J. Rohlf.
\newblock The comparison of dendrograms by objective methods.
\newblock \emph{Taxon}, pages 33--40, 1962.

\bibitem[Sud and Thelwall(2014)]{sud2014evaluating}
P.~Sud and M.~Thelwall.
\newblock Evaluating altmetrics.
\newblock \emph{Scientometrics}, 98\penalty0 (2):\penalty0 1131--1143, 2014.

\bibitem[Thelwall(2019)]{thelwall2019mendeley}
M.~Thelwall.
\newblock Mendeley reader counts for us computer science conference papers and
  journal articles.
\newblock \emph{Quantitative Science Studies}, pages 1--13, 2019.

\bibitem[Thelwall et~al.(2013)Thelwall, Haustein, Larivi{\`e}re, and
  Sugimoto]{thelwall2013altmetrics}
M.~Thelwall, S.~Haustein, V.~Larivi{\`e}re, and C.~R. Sugimoto.
\newblock Do altmetrics work? twitter and ten other social web services.
\newblock \emph{PloS one}, 8\penalty0 (5), 2013.

\bibitem[Tokuda et~al.(2020)Tokuda, Comin, and Costa]{tokuda2020revisiting}
E.~K. Tokuda, C.~H. Comin, and L.~d.~F. Costa.
\newblock Revisiting agglomerative clustering.
\newblock \emph{arXiv preprint arXiv:2005.07995}, 2020.

\bibitem[Waltman(2016)]{waltman2016review}
L.~Waltman.
\newblock A review of the literature on citation impact indicators.
\newblock \emph{Journal of Informetrics}, 10\penalty0 (2):\penalty0 365--391,
  2016.

\bibitem[Wang et~al.(2014)Wang, Liu, Fang, and Mao]{wang2014attention}
X.~Wang, C.~Liu, Z.~Fang, and W.~Mao.
\newblock From attention to citation, what and how does altmetrics work?
\newblock \emph{arXiv preprint arXiv:1409.4269}, 2014.

\bibitem[Watson(2009)]{watson2009comparing}
A.~B. Watson.
\newblock Comparing citations and downloads for individual articles at the
  journal of vision.
\newblock \emph{Journal of vision}, 9\penalty0 (4):\penalty0 i--i, 2009.

\end{thebibliography}

\newpage

\section*{Supplementary Information}

\renewcommand{\thefigure}{S\arabic{figure}}
\setcounter{figure}{0}

\begin{figure}[h]
    \centering
    \includegraphics[width=0.7\textwidth]{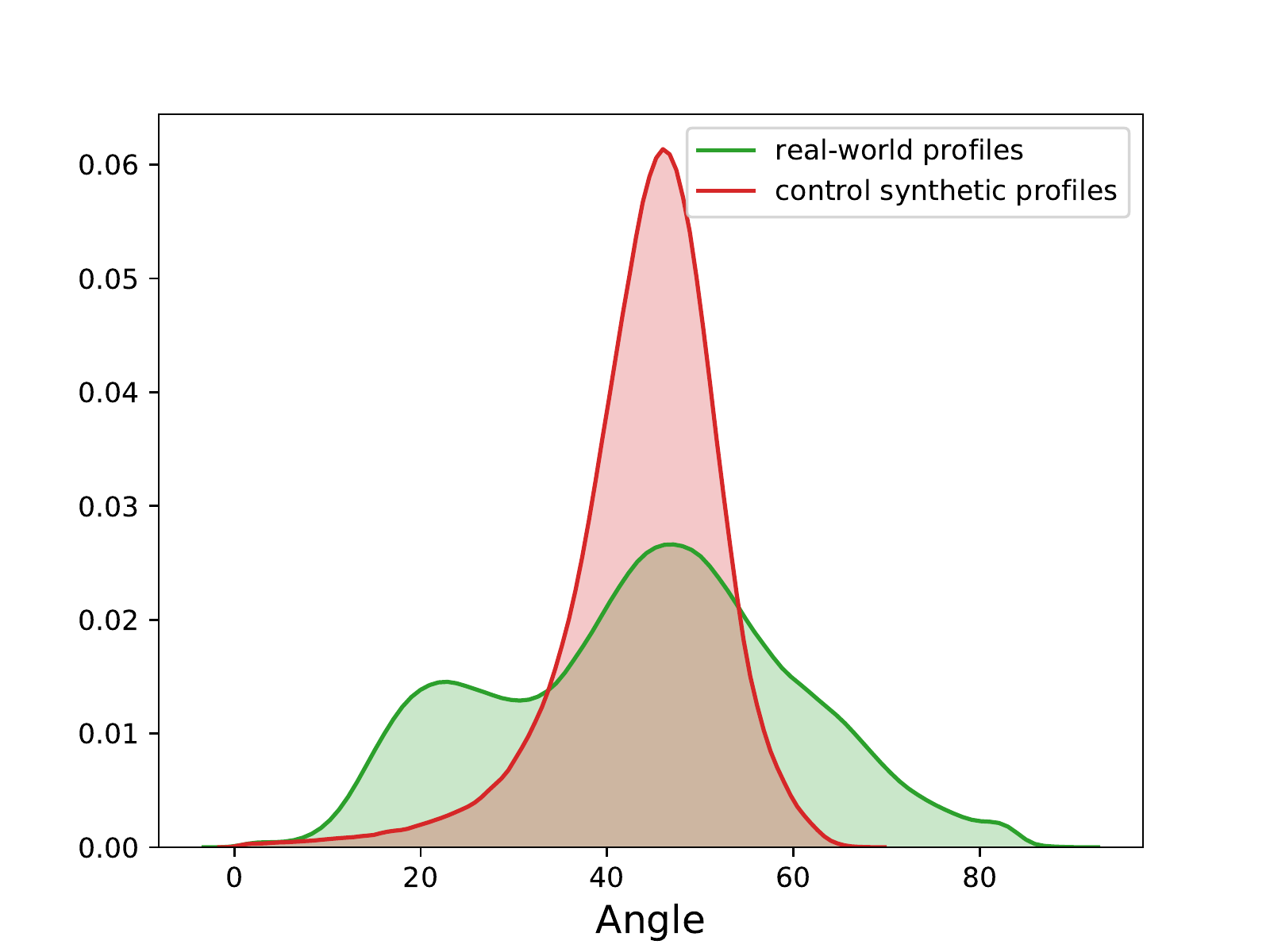}
    \caption{Angles distributions of real-world (green) and control (red) synthetic profiles. The distributions are clearly different, and a higher deviation in angles is observed for the real data.}
    \label{fig:angles}
\end{figure}

\begin{figure*}[h!]
\centering
\subfigure[Lifetime distribution for 2-segment profiles.]{\includegraphics[width=0.45\textwidth]{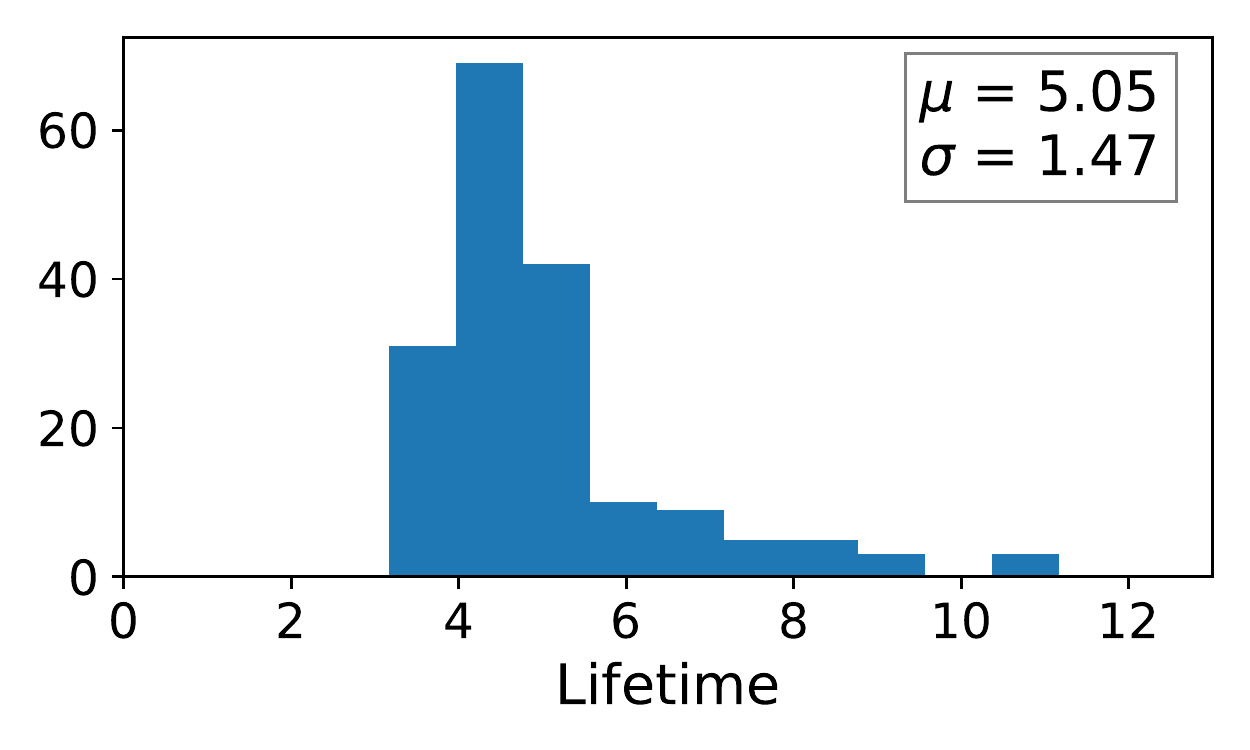}}
\subfigure[Lifetime distribution for 3-segment profiles.]{\includegraphics[width=0.45\textwidth]{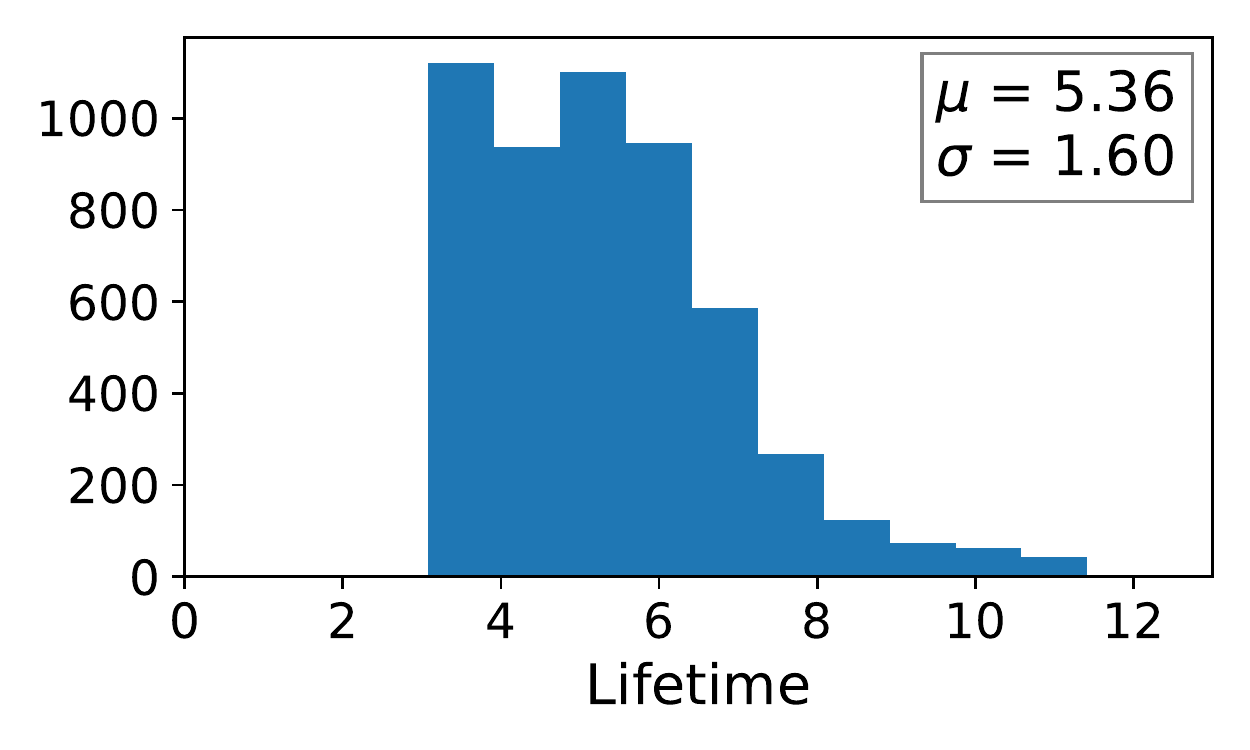}}
\subfigure[Lifetime distribution for 4-segment profiles.]{\includegraphics[width=0.45\textwidth]{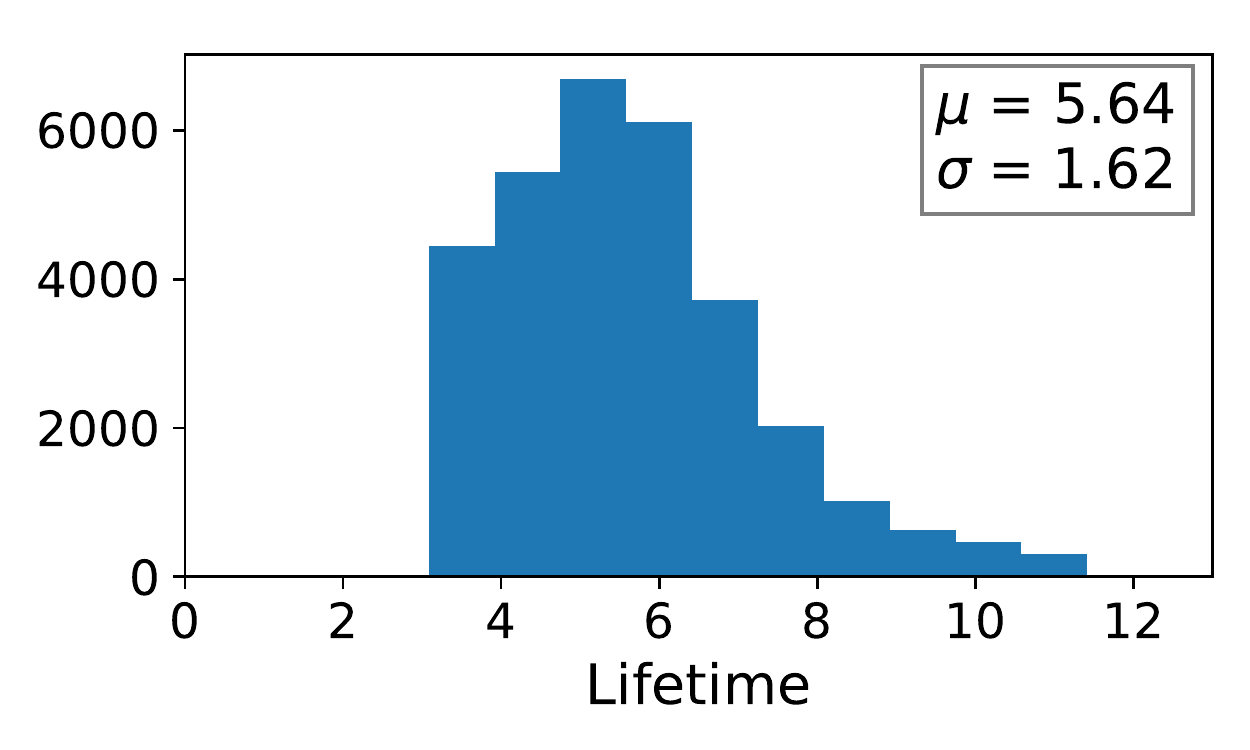}}
\subfigure[Lifetime distribution for 5-segment profiles.]{\includegraphics[width=0.45\textwidth]{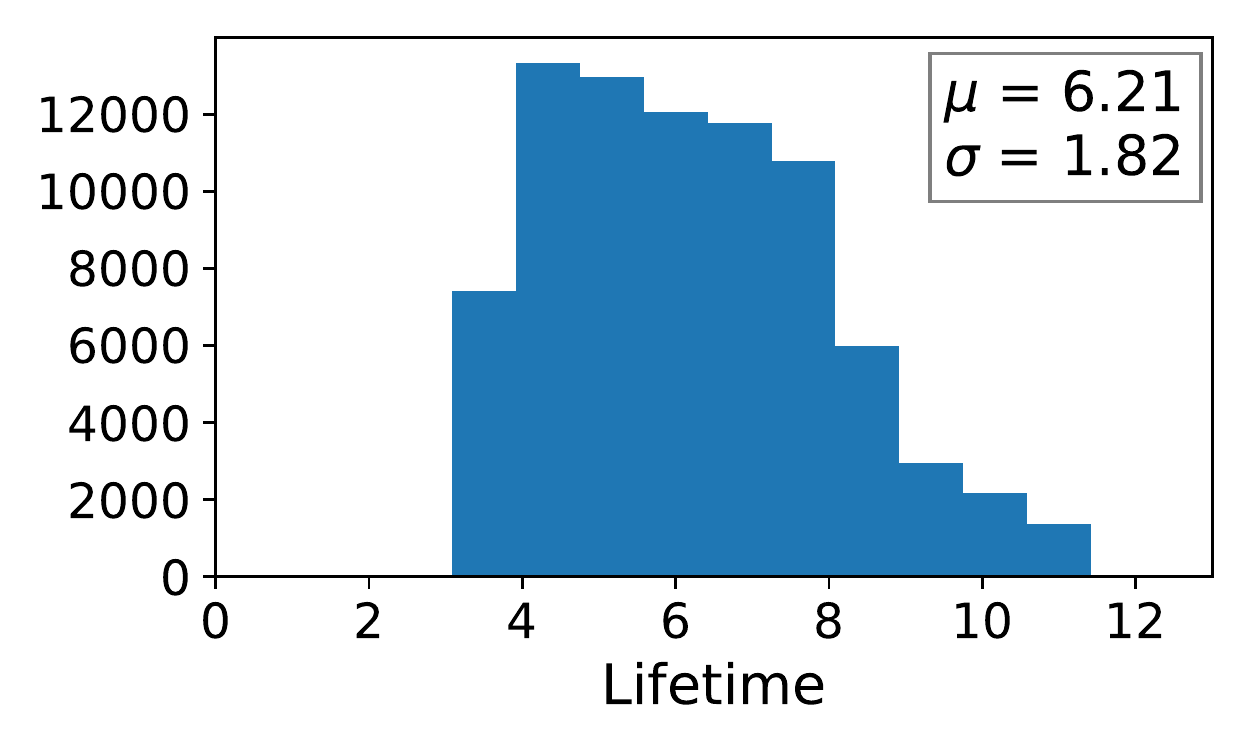}}
\caption{\label{fig:lifetimebysegments}Distributions of lifetime grouped by profiles with the same number of segments. Lifetime average and standard deviation are shown at the upper right corner.}
\end{figure*}

\begin{figure*}
\centering
\subfigure[$\rho$ = -0.08]{\includegraphics[width=6cm]{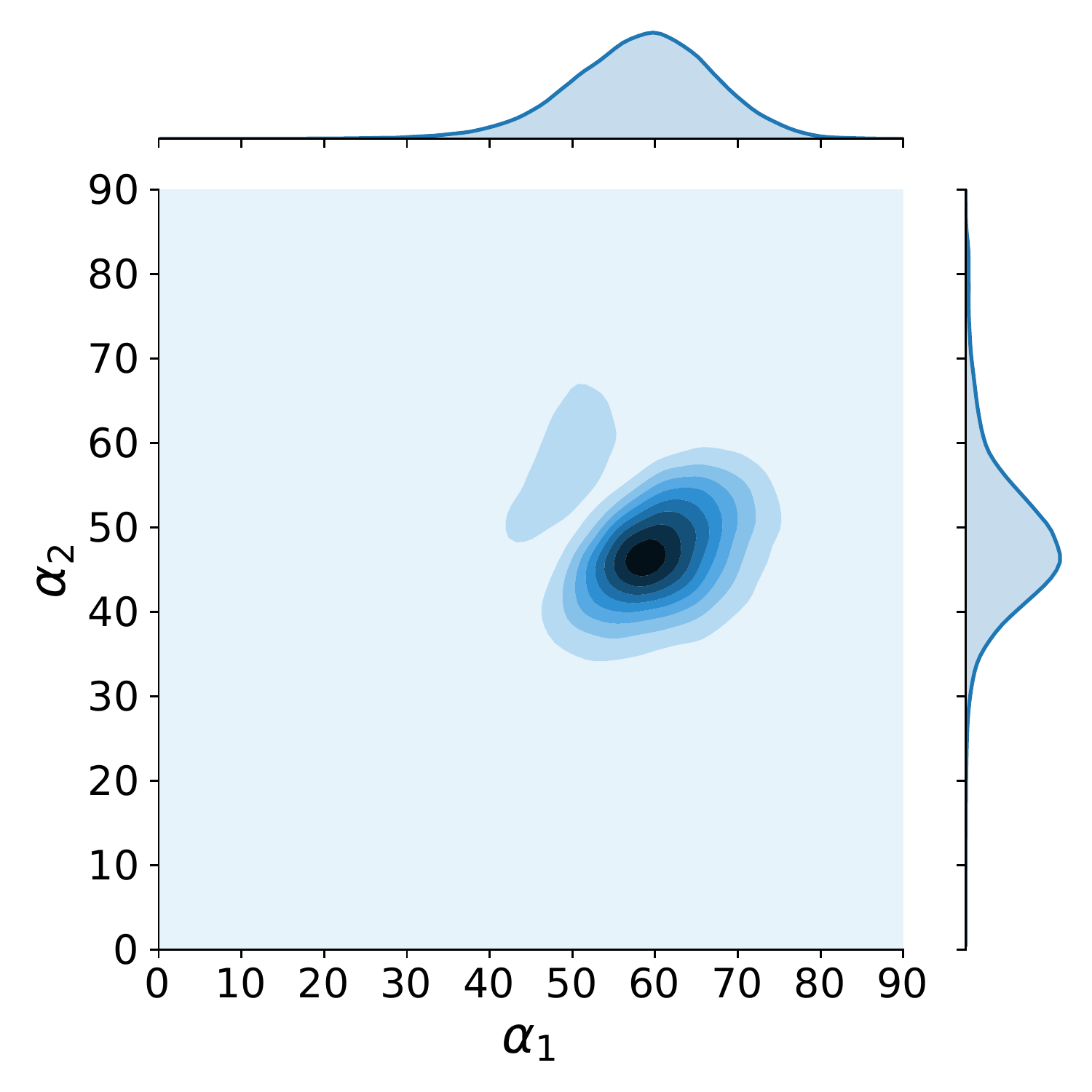}}
\subfigure[$\rho$ = -0.36]{\includegraphics[width=6cm]{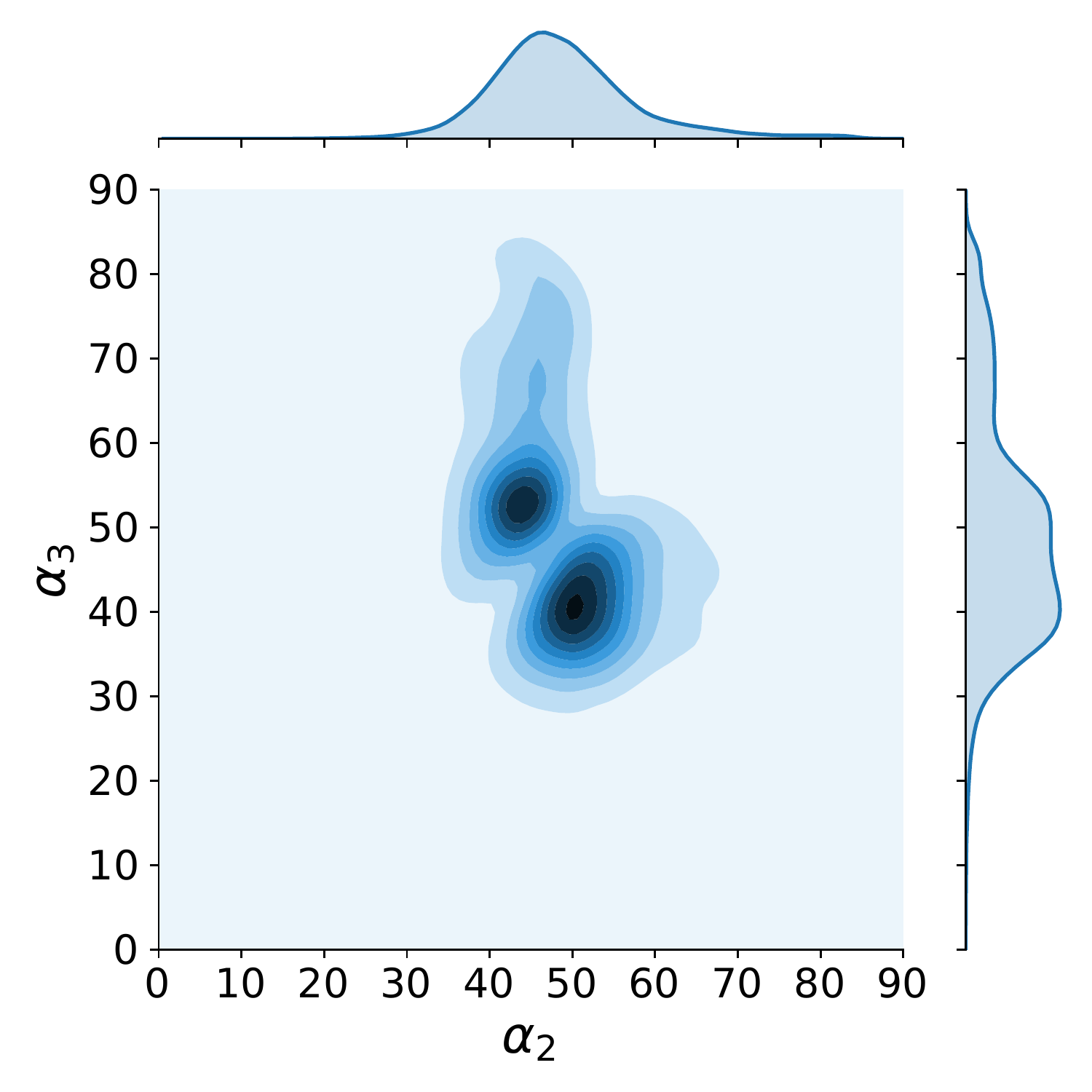}}
\subfigure[$\rho$ = -0.26]{\includegraphics[width=6cm]{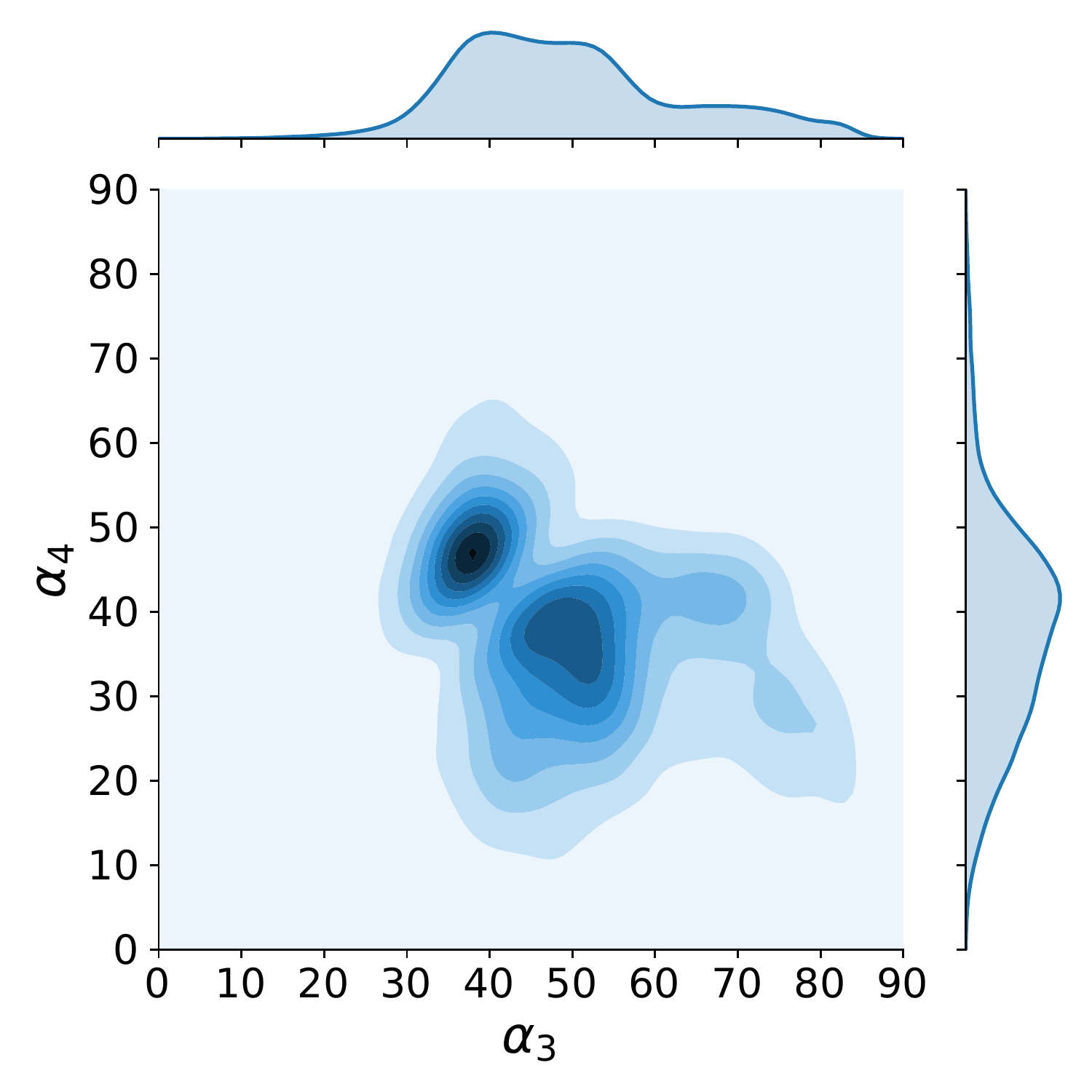}}
\subfigure[$\rho$ = 0.09]{\includegraphics[width=6cm]{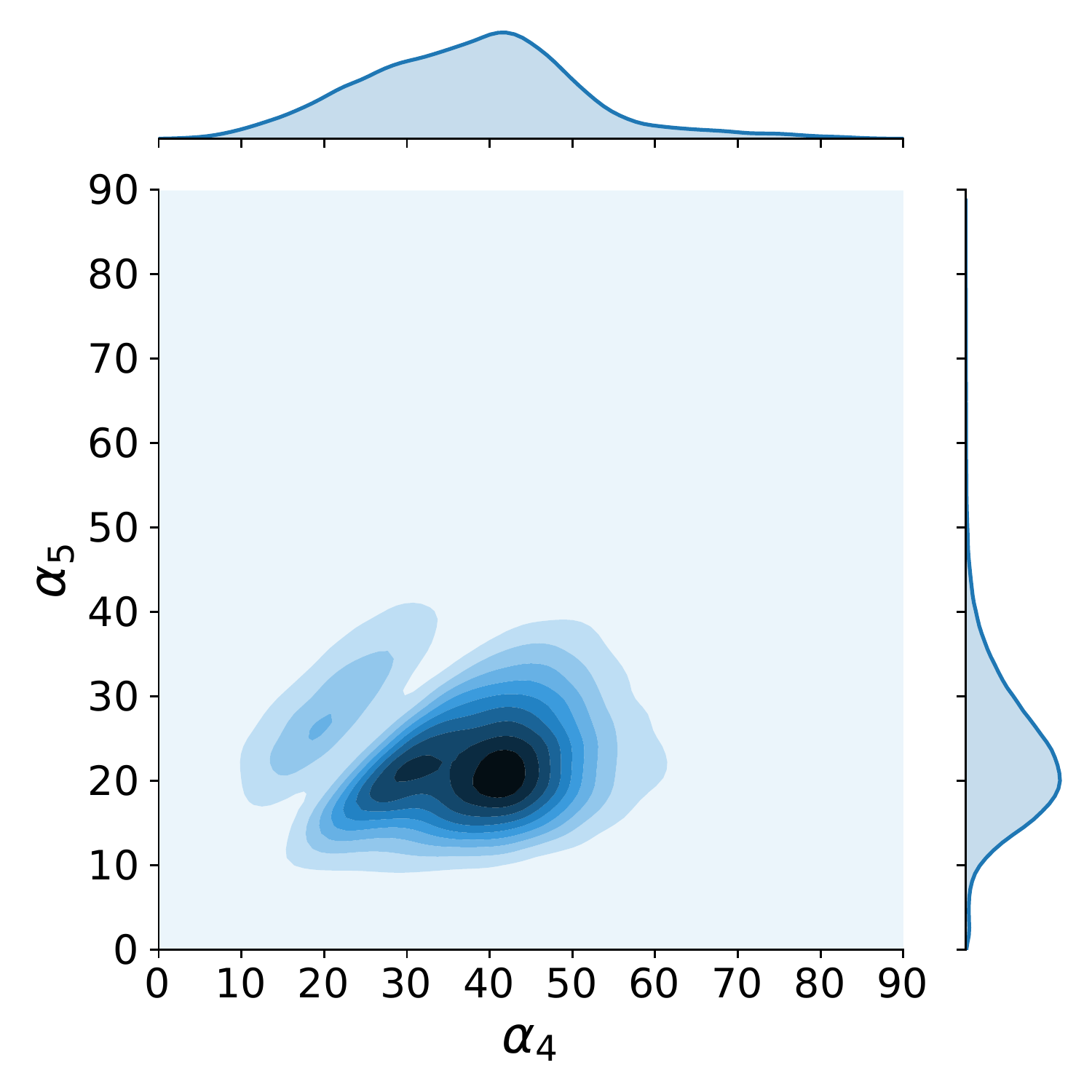}}
\caption{Joint density plots and corresponding marginal probabilities of $\alpha_i$ and $\alpha_{i+1}$ for profiles with five segments. $\rho$ is the Pearson correlation.}
\label{fig:jointalpha}
\end{figure*}

\begin{figure}
    \centering
    \includegraphics[width=0.87\textwidth]{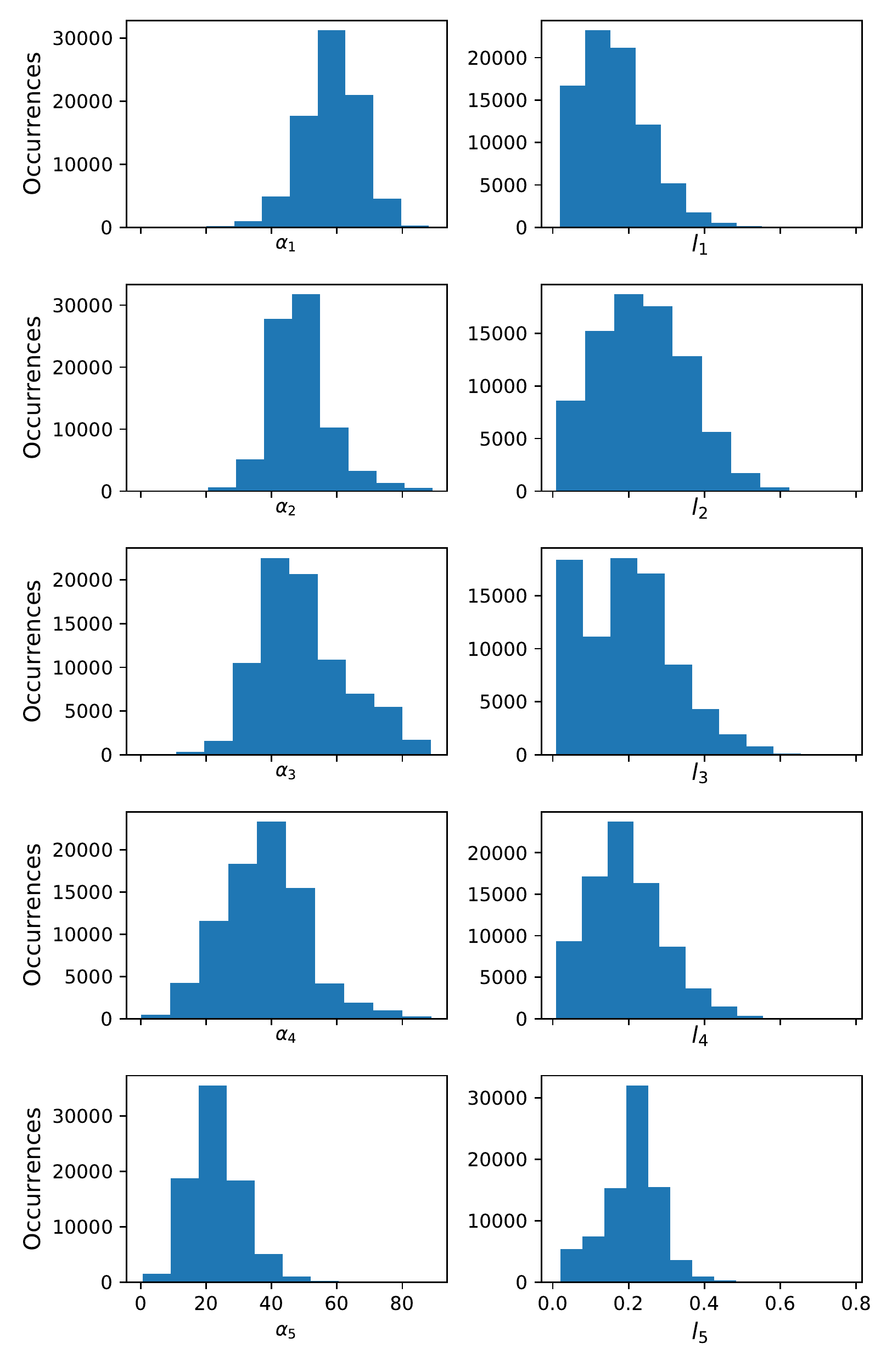}
    \caption{Distribution of the segment parameters among all the obtained curves. }
    \label{fig:histbyvar}
\end{figure}

%\begin{figure}[h]
%    \centering
%    \includegraphics[width=0.6\textwidth]{imgs/dist_intervals.pdf}
%    \caption{Intervals distributions of real-world and control synthetic profiles, respectively, green and red distributions.}
%    \label{fig:intervals}
%\end{figure}

% \begin{figure}
%     \centering
%     \includegraphics[width=0.6\textwidth]{imgs/lifetime_views.pdf}
%     \caption{\label{fig:bivariate}Bivariate distribution of lifetime and views. The Pearson correlation coefficient between these two variables is $\rho = 0.4087$.}
% \end{figure}

%\begin{figure}
%   \centering
%    \subfigure[Original data]{\includegraphics[width=0.4\textwidth]{imgs/vectors_diff_seaborn_orig.png}}
%    \subfigure[Synthetic data]{\includegraphics[width=0.4\textwidth]{imgs/vectors_diff_seaborn_syn.png}}
%    \caption{Caption}
%    \label{fig:my_label}
%\end{figure}

\end{document}